\documentclass[
  aps,
  pra,
  superscriptaddress,
  reprint
]{revtex4-2}

\usepackage[T1]{fontenc}
\usepackage[american]{babel}
\usepackage{newtxtext}

\usepackage[tbtags]{amsmath}
\usepackage{mathtools}
\usepackage{newtxmath}
\usepackage{subdepth}
\usepackage{physics}
\usepackage{mleftright}
\mleftright
\usepackage{bm}

\usepackage{scalerel}
\usepackage{graphicx}
\usepackage[HTML]{xcolor}

\usepackage{booktabs}
\usepackage{etoolbox}

\usepackage{orcidlink}
\usepackage{hyperref}
\usepackage[all]{hypcap}
\usepackage{adjustbox}
\definecolor{linkblue}{HTML}{2E3092}

\hypersetup{
  colorlinks = true,
  linkcolor  = linkblue,
  citecolor  = linkblue,
  urlcolor   = linkblue,
  filecolor  = linkblue
}

\newcommand\Dag{\stretchrel*{\dag}{X}}
\newcommand{\tupleentry}[1]{\makebox[1.8em][c]{$#1$}}
\newcommand{\tupleTwo}[2]{%
  (\tupleentry{#1},\tupleentry{#2})}
\newcommand{\tupleThree}[3]{%
  (\tupleentry{#1},\tupleentry{#2},\tupleentry{#3})}
  
\AtBeginDocument{%
  \setlength{\skip\footins}{18pt}%
}

\makeatletter
\patchcmd{\dch@set@one}{H_2}{H_b}
  {}{\errmessage{Subdepth patch 1 failed}}
\patchcmd{\dch@set@one}{H_2}{H_b}
  {}{\errmessage{Subdepth patch 2 failed}}
\patchcmd{\dch@set@one}{+\vrule}{\dagger\vrule}
  {}{\errmessage{Subdepth patch 3 failed}}
\makeatother

\usepackage{multirow}
\allowdisplaybreaks

\begin{document}
\title{Unified master equation for driven multilevel quantum systems:\texorpdfstring{\\}{}Derivation, dynamics, and spectral signatures}
\author{Sagarika Basak\,\orcidlink{0000-0003-2069-644X}}
\email[\!Contact author:~]{basak.sagarika@ou.edu}
\author{D.~Blume\,\orcidlink{0000-0001-7381-5698}}
\email[\!Contact author:~]{doerte.blume-1@ou.edu}
\affiliation{%
\adjustbox{max width=0.99\textwidth}{%
Homer L. Dodge Department of Physics and Astronomy,
\href{https://ror.org/02aqsxs83}{The University of Oklahoma},
440 W. Brooks Street, Norman, Oklahoma 73019, USA%
}}
\affiliation{%
\adjustbox{max width=\textwidth}{%
Center for Quantum Research and Technology,
\href{https://ror.org/02aqsxs83}{The University of Oklahoma},
440 W. Brooks Street, Norman, Oklahoma 73019, USA%
}}


\begin{abstract}
Drive-modified dissipation in multilevel quantum systems is not, in general, accurately captured by conventional master equations. We derive a unified master equation using Lewis--Riesenfeld dynamical invariants, accommodating arbitrary driving within Born--Markov theory. We demonstrate the applicability of the unified master equation to constant-Rabi-coupling (continuous-wave), periodically-modulated-Rabi-coupling, and nonperiodic (ramped and chirped) driving protocols. Fluorescence spectra of a periodically driven three-level $\Lambda$ system reveal signatures missed by conventional descriptions. We trace the limitations of the conventional descriptions to inaccurate bath-frequency sampling and the neglect of nonsecular dissipative couplings, and map out the parameter regimes that require a unified treatment. The fluorescence spectra provide experimentally accessible signatures, which serve as a spectroscopic guide to reliable modeling of dissipative processes that can inform quantum control, simulation, sensing, and information processing.
\end{abstract}




\maketitle
\section{\label{sec:level1}Introduction}
\vspace{-1em}
Quantum systems interact with their surroundings, exchanging energy and, in some settings, particles. The interactions with the environment or bath can produce dissipation and dephasing, modifying the otherwise unitary system dynamics. Understanding and controlling such effects is foundational to quantum science advances in sensing, information processing, and hybrid light--matter systems. A theoretical challenge lies in the consistent treatment of both the time-dependent coherent dynamics and the coupling to the bath~\cite{Dann2018,DiMeglio2024}. Common microscopic treatments based on the weak-coupling Born and memoryless Markov approximations result in the celebrated Redfield master equation~\cite{breuer2007}. A further secular approximation, which removes rapidly oscillating terms, can yield a Gorini--Kossakowski--Lindblad--Sudarshan (GKLS) generator that guarantees complete positivity~\cite{mccauley2020,Dodin2018}. However, for a system driven by external fields, evaluating the underlying propagator requires a time-ordered exponential, forcing conventional approaches to rely on further approximations that may yield inaccurate dynamics~\cite{shavit2019}. 

Two commonly used descriptions simplify the derivation of a master equation in different ways. In the ``laboratory-frame'' (LF) approach, the drive is omitted from the propagator used to construct the dissipators. This implies that the dissipative transition operators and rates are those of the undriven system~\cite{steck_quantum_atom_optics,Coleman2022}. The LF approach, which evaluates the bath spectral density $J(\omega)$ strictly at bare atomic transition frequencies~\cite{shavit2019,RoyChoudhury2016}, is appropriate when the drive has a negligible effect on the coupling operators over the bath correlation time. Adequate for weak driving in structureless baths, the LF approach fundamentally fails when driving alters dissipative transition channels or when the system is coupled to a structured bath such as a microwave photonic crystal~\cite{Harrington2019}, a surface-acoustic-wave device~\cite{Kitzman2023}, or a programmable trapped-ion simulator~\cite{Sun2025}. Alternatively, the conventional ``rotating-frame'' (RF) approach, defined for periodically driven systems, includes the drive in the dissipators by transforming to a rotating frame in which the system Hamiltonian is time independent and evaluating the system--bath interaction in the eigenstates of this Hamiltonian (i.e., in the dressed basis). Neglecting rapidly oscillating terms via a secular approximation, which requires strong driving, reduces the resulting dissipators to a tractable, time-independent GKLS form~\cite{Breuer1997,Kohler1997,shavit2019}. The conventional RF approach is thus suited for periodic high-frequency driving where dressed states are energetically separated (drive strength $\Omega_R \gg$ dissipation rate $\Gamma$)~\cite{shavit2019}. However, the conventional RF approach cannot accommodate arbitrary nonperiodic protocols or closed-loop multilevel coupling topologies~\cite{Paing2025}. Critically, in intermediate driving regimes, where drive-induced splittings are comparable to dissipative rates ($\Omega_R \sim \Gamma$), employing the secular approximation to enforce a tractable GKLS form discards essential dissipative dynamics~\cite{mccauley2020,Dodin2018,shavit2019}. While a generalized master equation has bridged this gap for driven two-level systems~\cite{shavit2019}, formulating an analogous description for multilevel systems under arbitrary driving remains an outstanding challenge.

We address this challenge by establishing a unified master-equation framework for driven dissipative $N$-level systems formulated through Lewis--Riesenfeld, or dynamical, invariants~\cite{wu2022,boubakour2025,basak2026}. Lewis--Riesenfeld invariants admit exact representations of nonadiabatic drive-induced evolution. As a consequence, our invariant-based master equation (IME) accommodates drives with constant Rabi coupling, periodically modulated Rabi coupling, and nonperiodic ramped or chirped modulation without relying on secular approximations or periodic Floquet conditions~\cite{Breuer1997,Kohler1997,DiMeglio2024}. Formulated in the eigenbasis of the dynamical invariant, the IME provides a unified master equation from which conventional LF models and a generalized RF master equation, derived in this work, emerge as limiting cases. Our generalized RF equation extends the conventional secular RF treatments~\cite{shavit2019} from the literature to time-dependent Rabi couplings and nonperiodic driving and recovers the conventional RF equation in the limit of periodic high-frequency driving. RF physics is obtained by discarding rapidly oscillating interaction-picture couplings with distinct dynamical frequencies between eigenstates of the invariant. LF physics is recovered by restricting bath spectral sampling to bare atomic frequencies and neglecting drive-induced modifications; reverting to the bare basis serves to recover the standard literature form~\cite{breuer2007}. Together, these considerations highlight which underlying approximations lead to the differences in the dynamics obtained from the three master equations. This interpretation is facilitated by the use of the eigenbasis of the dynamical invariant, which provides the shared representation of the three master equations. 

The remainder of this paper is organized as follows. 
Section~\ref{sec:thesystem} develops the unified master equation for an arbitrarily driven dissipative $N$-level system. 
Section~\ref{sec:threelevel} specializes the general framework to the driven three-level ($N=3$) $\Lambda$ system. 
Section~\ref{sec:protocols} examines the open-system dynamics under three qualitatively different driving protocols: constant-Rabi-coupling (continuous-wave), periodically-modulated-Rabi-coupling, and nonperiodic (ramped and chirped) driving.
Section~\ref{sec:spectrum} uses transition-resolved fluorescence spectra as an experimentally accessible diagnostic~\cite{Mollow1969,grove1977,Ng2022} to map the parameter regimes where conventional LF/RF descriptions fail, distinguishing discrepancies caused by bare-frequency bath sampling from those caused by discarding dissipative couplings by secular averaging. By delineating where conventional approximations suffice and where they fail, we aim to identify the simplest adequate master equation for a wide range of parameters.
Finally, Sec.~\ref{sec:summary} summarizes our main findings and presents an outlook. 
Technical derivations and additional background information are relegated to Appendices~\ref{sec:appendA}--\ref{sec:appendG}.

\section{\label{sec:thesystem}Unified generalized master equation}
We consider a driven $N$-level system coupled to a thermal bath. The total Hamiltonian is 
\begin{align}
    H(t)= H_s(t) +H_b+ H_I\,.
    \label{eqn:totalHamiltonian}
\end{align}
We set the energy of the state $N$ to zero and define $\omega^0_N = 0$. Throughout, we use units with $\hbar=1$. The driven $N$-level system is governed by the system Hamiltonian
\begin{align}
    H_s(t) = -\smashoperator{\sum\limits_{\eta<N}} \omega^0_{\eta} \sigma_{\eta,\eta}\nonumber 
    +& \smashoperator[l]{\sum\limits_{\eta, \upsilon>\eta}} \Bigg\{\dfrac{1}{2}\left(e^{+i\omega^d_{\eta,\upsilon}t} + e^{-i\omega^d_{\eta,\upsilon}t}\right) \nonumber\\[-6pt]
    &\;\;\times\left[\Omega_{\eta,\upsilon}(t) \sigma^{\Dag}_{\eta,\upsilon}+\Omega^*_{\eta,\upsilon}(t) \sigma_{\eta,\upsilon}\right]\Bigg\}\,,
    \label{eqn:generalsystemHamiltonian}
\end{align}
where $\sigma_{\eta,\upsilon}=\ket{\eta}\bra{\upsilon}$. The states $\ket{\eta}$ are the eigenstates of the undriven system Hamiltonian and are referred to as the \textit{bare states}. $\omega^0_{\eta}$ is the transition frequency for the $(N{\leftrightarrow}\eta)$ transition, $\Omega_{\eta,\upsilon}(t)$ is the generally complex time-dependent Rabi coupling, and $\omega^d_{\eta,\upsilon}$ is the time-independent frequency of the field that drives the $(\upsilon{\leftrightarrow}\eta)$ transition. The exponential factors in Eq.~\eqref{eqn:generalsystemHamiltonian} describe the periodic oscillations at the drive frequencies $\omega^d_{\eta,\upsilon}$, whereas the time dependence of $\Omega_{\eta,\upsilon}(t)$ allows for an arbitrary modulation of the drive envelope. Equation~\eqref{eqn:generalsystemHamiltonian} encompasses constant-Rabi-coupling (continuous-wave), periodically-modulated-Rabi-coupling, and nonperiodic (ramped and chirped) driving protocols. 

The thermal bath is modeled by a set of harmonic oscillators, 
\begin{align}
    H_b = \sum\limits_k \omega_k b^{\Dag}_kb_k \,.
    \label{eqn:bathHamiltonian}
\end{align}
Here, $b^{\Dag}_k$ creates and $b_k$ annihilates a bath excitation in mode $k$ with frequency $\omega_k$ ($\omega_k$ is real). The system--bath interaction Hamiltonian is given by
\begin{align}
    H_I = \sum\limits_{\eta,\,\upsilon} \sigma_{\eta,\upsilon}\otimes\left[\sum\limits_k g_{\eta,\,\upsilon,\,k}\left(b_k+b^{\Dag}_k\right)\right]\,,
    \label{eqn:systembathintHamiltonian}
\end{align}
with $g_{\eta,\upsilon,k} = e^{i\varphi_{\eta,\upsilon}}\abs*{g_{\eta,\upsilon,k}}$ denoting the system--bath coupling coefficients, subject to $g_{\upsilon,\eta,k}=g_{\eta,\upsilon,k}^*$ so that $H_I$ is Hermitian. This is imposed by setting $\varphi_{\eta,\upsilon} = - \varphi_{\upsilon,\eta}$ and $\abs*{g_{\eta,\upsilon,k}} = \abs*{g_{\upsilon,\eta,k}}$. Equation~\eqref{eqn:systembathintHamiltonian} includes off-diagonal (transverse) couplings, which induce transitions between bare states, and diagonal (longitudinal) couplings, which generate dephasing in the bare basis. For each bath mode $k$, the Hermitian system operator $\sum_{\eta,\upsilon}g_{\eta,\upsilon,k}\sigma_{\eta,\upsilon}$ can be expanded in the identity and the generators of $\mathrm{SU}(N)$. 

We assume a continuum of bath modes and describe the system--bath couplings through the spectral-density functions
\begin{align}
    J_{\eta,\upsilon}(\omega) &= \sum_k \abs*{g_{\eta,\upsilon,k}}^2 \delta\left(\omega-\omega_k\right)
    \label{eqn:continuummodes1}\\[-6pt]
\intertext{and}
\noalign{\vskip -6pt}
    J_{\eta,\upsilon,\eta',\upsilon'}(\omega) &= \sum_k g_{\eta,\upsilon,k}g^*_{\eta',\upsilon',k}\delta\left(\omega-\omega_k\right)\,,
    \label{eqn:continuummodes2}
\end{align}
where $J_{\eta,\upsilon}(\omega) = J_{\eta,\upsilon,\eta,\upsilon}(\omega)$ is the diagonal spectral density for the transition $(\eta,\upsilon)$. The off-diagonal (or cross) elements $J_{\eta,\upsilon,\eta',\upsilon'}(\omega)$ with $(\eta,\upsilon)\neq(\eta',\upsilon')$ capture bath-mediated correlations between distinct bare transitions through the coupling to the bath. At this stage, no particular functional form of the spectral density is assumed, i.e., the distribution of the coupling coefficients $g_{\eta,\upsilon,k}$ is left unspecified; the subsequent master-equation construction therefore applies to, e.g., Ohmic, sub-Ohmic, super-Ohmic, or structured spectral densities. A specific example will be presented in a later section.

In what follows, we assume that the bath is thermal. Starting from the Bloch--Redfield master equation (see Appendix~\ref{sec:appendA}), we find that the reduced dynamics of the driven open quantum system within the Born--Markov approximation can be described by the master equation 
\begin{widetext}
\vspace*{-1.5em}
\begin{align}
    &\dfrac{d}{dt}\rho(t) = -i\comm{H_{s}(t)}{\rho(t)}+ \smashoperator{\sum\limits_{m,n,m',n'}}\Gamma_{m,n,m',n'}(t)\left[F_{m,n}(t)\rho(t)F_{m',n'}^{\Dag}(t) -F_{m',n'}^{\Dag}(t)F_{m,n}(t)\rho(t)\right] + \text{H.c.} 
     \label{eqn:unifiedgenME}
\end{align}
\end{widetext}
regardless of whether the derivation is carried out using the IME, LF, or generalized RF approach. In Eq.~\eqref{eqn:unifiedgenME}, $\rho(t)$ is the reduced density matrix of the system and
\begin{align}
F_{m,n}(t) = \ket{\mu_m(t)}\bra{\mu_n(t)} 
\end{align}
are the jump operators in the eigenbasis of the dynamical invariant \cite{wu2022,boubakour2025}
\begin{align}
    I(t) = \sum_n \lambda_n \ket{\mu_n(t)}\bra{\mu_n(t)}\,.
    \label{eqn:dynamicalInv}
\end{align}
Throughout, the eigenvalues $\lambda_n$ are assumed to be real and time independent, and the eigenstates $\ket{\mu_n(t)}$ form an orthonormal basis. The eigenvalues $\lambda_n$ are nondegenerate for the cases considered in this work. The invariant satisfies \cite{wu2022,boubakour2025}
\begin{align}
    \dfrac{\partial }{\partial t}I(t) = i\comm{I(t)}{H_{s}(t)}\,.
    \label{eqn:invariantcondition}
\end{align}
We refer to Eq.~\eqref{eqn:unifiedgenME} as the unified master equation since it provides a common functional form that applies to the IME, LF, and generalized RF master equations. Specifically, the three frameworks differ in the expressions for the dissipation coefficients $\Gamma_{m,n,m',n'}(t)$. It should be noted that LF and conventional RF master equations are typically not expressed in terms of jump operators that utilize the eigenstates of the invariant but in terms of jump operators that utilize the bare or dressed states (dressed states are the eigenstates of the system Hamiltonian in the rotating frame). The connection is discussed in Appendices~\ref{sec:appendD} and \ref{sec:appendE}.

To determine the coefficients $\Gamma_{m,n,m',n'}(t)$, we decompose the matrix elements of the atomic transition operator $\sigma_{\eta,\upsilon}$, expressed in the invariant eigenbasis, into a time-dependent phase $\varsigma_{m,n,\eta,\upsilon}(t)$ and a complex transition amplitude $\xi_{m,n,\eta,\upsilon}(t)$,
\begin{align}
    \mel{\mu_{m}(t)}{\sigma_{\eta,\upsilon}}{\mu_n(t)} 
    = \xi_{m,n,\eta,\upsilon}(t)\, e^{i\varsigma_{m,n,\eta,\upsilon}(t)}\,, 
    \label{eqn:matrixelements}
\end{align}
where we have absorbed the constant phase $\varsigma^0_{m,n,\eta,\upsilon}$ into $\xi_{m,n,\eta,\upsilon}(t)$,
\begin{align}
    \xi_{m,n,\eta,\upsilon}(t) 
    = \abs{\mel{\mu_{m}(t)}{\sigma_{\eta,\upsilon}}{\mu_n(t)}} e^{i\varsigma^0_{m,n,\eta,\upsilon}}\,.
    \label{eqn:amplitude}
\end{align}
We define the total phase as
\begin{align}
    \theta_{m,n,\eta,\upsilon}(t) = \alpha_n(t)-\alpha_m(t) +\varsigma_{m,n,\eta,\upsilon}(t)\,,
    \label{eqn:phases}
\end{align}
where the Lewis--Riesenfeld phases $\alpha_n(t)$ are given by
\begin{align}
    \alpha_n(t) = \textstyle\int\limits_{0}^{t}\expval{i \dfrac{\partial}{\partial \tau}-H_s(\tau)}{\mu_n(\tau)}d\tau\,.
    \label{eqn:LRIphases}
\end{align}
Expanding $\theta_{m,n,\eta,\upsilon}(t-\tau)$ around $t$,  
\begin{align}
   \theta_{m,n,\eta,\upsilon}(t-\tau) =\theta_{m,n,\eta,\upsilon}(t) -\omega_{m,n,\eta,\upsilon}(t,\tau)\tau\,,
    \label{eqn:thetaexpand}
\end{align}
defines the instantaneous frequencies 
\begin{align}
    \omega_{m,n,\eta,\upsilon}(t,\tau) = \sum_{r>0}^{\infty} \dfrac{(-\tau)^{r-1}}{r!} \dfrac{d^r}{dt^r} \theta_{m,n,\eta,\upsilon}(t)\,.
    \label{eqn:instfreq}
\end{align}

\begin{figure*}
\centering
\includegraphics{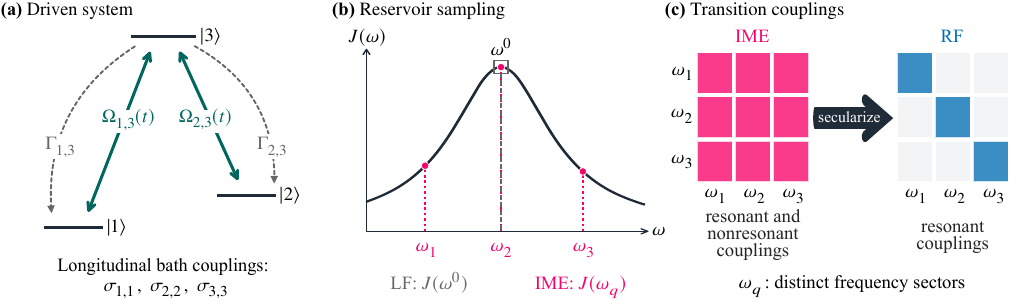}
\caption{Illustration of drive-modified dissipation.
(a) Driven three-level $\Lambda$ system with transverse and longitudinal system--bath couplings.
(b) Bath sampling. The IME approach samples the bare and drive-modified transition frequencies while the LF approach only samples the bare transition frequencies. 
(c) Transition couplings. The IME approach accounts for couplings between all frequency sectors while the generalized RF approach only retains couplings between resonant frequency sectors. In this stationary illustration, each frequency sector groups degenerate transitions; non-gray color indicates the couplings retained and not their magnitudes.}
\label{fig:framework}
\end{figure*}

\noindent With these definitions in place, we are ready to provide explicit expressions for the coefficients $\Gamma_{m,n,m',n'}(t)$. The dissipation coefficients for the IME, LF, and generalized RF approaches are derived in Appendices~\ref{sec:appendB}, \ref{sec:appendD}, and \ref{sec:appendE}, respectively. They are given by 
\begin{widetext}
\vspace{-1.5em}
\begin{flalign}
    \Gamma_{m,n,m',n'}(t) \underset{\text{IME}}{=}& \smashoperator[r]{\sum\limits_{\substack{\eta\hphantom{'},\,\upsilon\hphantom{'}\\ \eta',\,\upsilon'}}}  e^{i\left[\varsigma_{\vphantom{m',n'}m,n,\vphantom{\eta',\upsilon'}\eta,\upsilon}(t)-\varsigma_{m',n',\eta',\upsilon'}(t)\right]}[\xi_{m',n',\eta',\upsilon'}(t)]^* \int\limits_0^{\infty}\! \bm{\Bigg(} d\tau \xi_{m,n,\eta,\upsilon}(t-\tau)  e^{-i\omega_{m,n,\eta,\upsilon}(t,\tau)\tau} \nonumber \\[-9pt]
    &\qquad\qquad\qquad\qquad\; \times\! \left\{\textstyle\int\limits_0^{\infty}\! d\omega J_{\eta,\upsilon,\eta',\upsilon'}\left(\omega\right)\left[n_b(\omega)e^{+i\omega \tau } + \left[1+n_b(\omega)\right]e^{-i\omega \tau }\right]\right\}\bm{\Bigg)}\,,&&
    \label{eqn:disscoefficient-IME}
\end{flalign}
\begin{flalign}
    \Gamma_{m,n,m',n'}(t) \underset{\text{RF}}{=}& \bm{\Bigg[}\smashoperator[r]{\sum\limits_{\substack{\eta\hphantom{'},\,\upsilon\hphantom{'}\\ \eta',\,\upsilon'}}} [\xi_{m',n',\eta',\upsilon'}(t)]^* \int\limits_0^{\infty}\! \bm{\Bigg(} d\tau \xi_{m,n,\eta,\upsilon}(t-\tau)  e^{-i\omega_{m,n,\eta,\upsilon}(t,\tau)\tau} \nonumber\\[-9pt]
    &\qquad\qquad\qquad\qquad\;\;\; \times\! \left\{\textstyle\int\limits_0^{\infty}\! d\omega J_{\eta,\upsilon,\eta',\upsilon'}(\omega)\left[n_b(\omega)e^{+i\omega \tau } + \left[1+n_b(\omega)\right]e^{-i\omega \tau }\right]\right\}\bm{\Bigg)}\bm{\Bigg]}\Bigg\rvert_{\theta_{\vphantom{m',n'}m,n,\vphantom{\eta',\upsilon'}\eta,\upsilon}\!(t)=\theta_{m',n',\eta',\upsilon'}\!(t)}\!,&&
    \label{eqn:disscoefficient-RF}\raisetag{25pt}
\end{flalign}
and
\begin{flalign}
    \Gamma_{m,n,m',n'}(t) \underset{\text{LF}}{=}& \smashoperator[r]{\sum\limits_{\substack{\eta\hphantom{'},\,\upsilon\hphantom{'}\\ \eta',\,\upsilon'}}} e^{i\left[\varsigma_{\vphantom{m',n'}m,n,\vphantom{\eta',\upsilon'}\eta,\upsilon}(t)-\varsigma_{m',n',\eta',\upsilon'}(t)\right]}  [\xi_{m',n',\eta',\upsilon'}(t)]^*\xi_{m,n,\eta,\upsilon}(t) \int\limits_0^{\infty}\! \bm{\Bigg(} d\tau e^{+i\omega^0_{\eta,\upsilon} \tau} \nonumber \\[-9pt]
    &\qquad\qquad\qquad\qquad\;\;\; \times\! \left\{\textstyle\int\limits_0^{\infty}\! d\omega J_{\eta,\upsilon,\eta',\upsilon'}\left(\omega\right)\left[n_b(\omega)e^{+i\omega \tau } + \left[1+n_b(\omega)\right]e^{-i\omega \tau }\right]\right\}\bm{\Bigg)}\,,&&
    \label{eqn:disscoefficient-LF}
\end{flalign}
where $n_b(\omega) = \left(e^{\omega/k_{\text{B}}T_b}-1\right)^{-1}$ is the Planck distribution and $T_b$ denotes the temperature of the thermal bath. 
\end{widetext}
Building on previous invariant-based constructions~\cite{wu2022,Dann2018,Dann2019,boubakour2025}, the dissipative terms are formulated in the invariant eigenbasis. As shown in Appendix~\ref{sec:appendB}, the use of the eigenstates of the invariant greatly simplifies the derivation. Importantly, though, the LF master equation can alternatively be derived, as is well known, using the bare atomic states~\cite{breuer2007,rivas2012}. The generalized RF master equation presented above is, as discussed in more detail below, more general than the conventional RF master equations considered in the literature.

The unified master equation provides a systematic description of driven open quantum systems with constant-Rabi-coupling, periodically-modulated-Rabi-coupling, and nonperiodic (ramped and chirped) driving protocols within the Born--Markov approximation. Although the IME, generalized RF, and LF treatments share the common structure of Eq.~\eqref{eqn:unifiedgenME}, they differ in which system transitions are coupled to the bath as well as how strong the coupling is [see Figs.~\ref{fig:framework}(b) and \ref{fig:framework}(c) for illustrations]. The IME incorporates the full time-dependent complex transition amplitudes and drive-modified frequencies. Accounting for these effects is facilitated by the use of the dynamical-invariant basis. The LF treatment approximates the dissipators by neglecting the coherent drive on the time scale of the bath correlation time. Consequently, the bath is sampled at the bare atomic frequencies $\omega^0_{\eta,\upsilon}$ and the complex transition amplitudes $[\xi_{m',n',\eta',\upsilon'}(t)]^*$ and $\xi_{m,n,\eta,\upsilon}(t)$ [see Eq.~\eqref{eqn:disscoefficient-LF}] are both evaluated at the same time $t$ rather than at $t$ and $t-\tau$ as in the IME and generalized RF treatments. Conversely, the generalized RF treatment accounts for the drive-modified transition structure and memory, but drops nonsecular dissipative couplings, i.e., the coupling between transitions with distinct dynamical frequencies [see Fig.~\ref{fig:framework}(c) for an illustration]. The differences in the dissipation coefficients provide the basis for the comparative analyses presented in the following sections. 

In the absence of any driving, the IME and LF descriptions coincide, and application of the secular approximation to distinct bare transition frequencies yields the standard GKLS master equation~\cite{breuer2007,mccauley2020}. For a periodically driven two-level system, Appendix~\ref{sec:appendC} establishes how the IME reproduces the generalized master equation introduced in Ref.~\cite{shavit2019}. The fact that the IME master equation reproduces the generalized master equation for $N = 2$ highlights the utility of the IME approach as a unifying framework. The conventional RF construction~\cite{shavit2019} assumes periodic driving so that moving to a frame that rotates at the drive frequencies, together with the rotating-wave approximation, yields a time-independent Hamiltonian. The derivation furthermore makes a secular approximation, which requires that the separation between distinct dressed-state transition frequencies greatly exceeds the dissipation rates. Our generalized RF construction, in contrast, allows for time-dependent Rabi couplings and nonperiodic protocols. The derivation of our generalized RF approach and its connection to the conventional RF master equation are presented in Appendix~\ref{sec:appendE}. When the drive is periodic and the transformation renders the Hamiltonian effectively time independent under the rotating-wave approximation~\cite{Paing2025}, the generalized RF equation reduces to the conventional RF master equation~\cite{shavit2019}. In contrast to Floquet master equations, which require strict periodicity~\cite{Breuer1997,Kohler1997}, and adiabatic master equations, which rely on slow driving~\cite{DiMeglio2024}, the IME imposes neither restriction. 

Importantly, the reduced system dynamics described by the IME approach are, just as many literature approaches, subject to the Born--Markov approximation. Because the IME makes no dynamical approximations beyond the Born--Markov assumptions, it retains the structure of the Redfield master equation and cannot, in general, be written in GKLS form. In contrast, the LF and generalized RF master equations can be written in GKLS (Lindblad) form (see Appendices~\ref{sec:appendD} and \ref{sec:appendE}) and are therefore characterized by complete positivity~\cite{fernandezdelapradilla2024}. The positivity is enforced by invoking secular approximations on the bare transitions in the LF approach and the rotating-wave and secular approximations on the dressed transitions in the RF approach. Consequently, potential loss of positivity in the IME stems from the underlying structure of the Redfield master equation itself. Rather than indicating a flaw in the IME, the emergence of negativity serves as a diagnostic signature that the chosen parameters, e.g., large dissipation rates, push beyond the validity regime of the weak-coupling Born--Markov approximation. In contrast, the secular approximations that are made when deriving the LF and generalized RF equations enforce positivity mathematically~\cite{fernandezdelapradilla2024}, which masks the physical breakdown rather than correcting it and introduces inaccurate system dynamics whenever the weak-coupling Born--Markov approximation is violated. Finally, we note that while a given master equation can always be rewritten in another basis, writing the LF and generalized RF equations as limiting cases of the IME places them within a single, common representation, transparently delineating the source and magnitude of differences that arise from the specific frame in which the secular and rotating-wave approximations are being made.

\section{\label{sec:threelevel}Driven three-level system}
In this and the following sections, we focus on a three-level system in the $\Lambda$ configuration, with the $(3{\leftrightarrow}1)$ and $(3{\leftrightarrow}2)$ transitions driven [see Fig.~\ref{fig:framework}(a)]. Assuming
\begin{align}
\omega^d_{\eta,3} \gg \sqrt{\sum_{\eta} \left(\Omega^2_{\eta,3}(t)+\Delta^2_{\eta}\right)}\,, 
\end{align}
where $\Delta_{\eta} = \omega^0_{\eta} - \omega^d_{\eta,3}$, and rotating to a frame in which the states oscillate at the drive frequencies using
\begin{align}
U(t) = \exp\left(-i\sum\limits_{\eta = 1,2}\omega^d_{\eta,3} \sigma_{\eta,\eta}t\right)\,,
\label{eqn:Urot}
\end{align}
the Hamiltonian, after applying the rotating-wave approximation, becomes 
\begin{align}
    H^R_s(t)=\sum\limits_{\eta = 1,2}\left[-\Delta_{\eta} \sigma_{\eta,\eta}+\dfrac{1}{2}\left(\Omega_{\eta,3}(t)\sigma^{\Dag}_{\eta,3}+\Omega^*_{\eta,3}(t)\sigma_{\eta,3}\right)\right]\,. 
    \label{eqn:3levelHamRF}
\end{align}
It can be seen that switching to the rotating frame and applying the rotating-wave approximation allows us to remove fast oscillations and systematically organize terms order by order across the relevant energy scales. We emphasize that this is merely a convenient analytical and numerical tool rather than a fundamental requirement for the derivation of the master equation. Specifically, it allows us to systematically isolate energy scales by, e.g., discarding fast oscillations that would otherwise need to be treated numerically, while retaining the time dependence of the Rabi couplings, which are assumed to vary much more slowly. For the driving protocols considered below, we take equal detunings, i.e., we set $\Delta_1=\Delta_2=\Delta$. We define the instantaneous generalized Rabi coupling $\Omega_T(t)=\sqrt{\Delta^2+\abs{\Omega_{1,3}(t)}^2+\abs{\Omega_{2,3}(t)}^2}$.

Rotating the frame transforms the system--bath interaction Hamiltonian according to $U^{\Dag}\!(t)H_I U(t)$. For the interaction Hamiltonian in Eq.~\eqref{eqn:systembathintHamiltonian} with $N=3$, the rotated system--bath interaction Hamiltonian takes the form
\begin{align}
    H^R_I(t) =\sum\limits_{\eta,\,\upsilon} e^{-i\omega^d_{\eta,\upsilon}t}\sigma_{\eta,\upsilon}\otimes\left[ \sum\limits_k g_{\eta,\,\upsilon,\,k} \left(b_k+b^{\Dag}_k\right)\right]\,,
    \label{eqn:systembathintHamiltonianrot}
\end{align}
where $\omega^d_{1,2} = \omega^d_{1,3} - \omega^d_{2,3}$,  $\omega^d_{\upsilon,\eta} = -\omega^d_{\eta,\upsilon}$, and $\omega^d_{\eta,\eta} = 0$. Using the rotated system operators in Eq.~\eqref{eqn:systembathintHamiltonianrot}, the eigenbasis and the equations of motion of the dynamical invariant defined in Appendix~\ref{sec:appendF}, as well as the Lewis--Riesenfeld phases [Eq.~\eqref{eqn:LRIphases}], we obtain the matrix elements [Eq.~\eqref{eqn:matrixelements}], phases [Eq.~\eqref{eqn:phases}], and instantaneous frequencies [Eq.~\eqref{eqn:instfreq}].

For the applications considered in this paper, we evaluate the dissipation coefficients by assuming that the magnitudes of the transition amplitudes $\xi_{m,n,\eta,\upsilon}(t)$ vary negligibly over the bath correlation time and that the total phases $\theta_{m,n,\eta,\upsilon}(t)$ vary approximately linearly over the same interval. We therefore use
\begin{align} 
\xi_{m,n,\eta,\upsilon}(t-\tau) &\approx \xi_{m,n,\eta,\upsilon}(t)\label{eqn:local_drive_approximations1} 
\end{align}
and
\begin{align} 
\theta_{m,n,\eta,\upsilon}(t-\tau) &\approx \theta_{m,n,\eta,\upsilon}(t) -\dfrac{d{\theta}_{m,n,\eta,\upsilon}(t)}{dt}\tau\,. \label{eqn:local_drive_approximations2} 
\end{align}
For constant Rabi couplings, the chosen rotating-frame invariant eigenbasis is time independent and the total phases are linear in time, making the replacements in Eqs.~\eqref{eqn:local_drive_approximations1} and \eqref{eqn:local_drive_approximations2} exact. The initial state for all the cases considered below is $\ket{1}\bra{1}$. For the spectral analysis presented in Sec.~\ref{sec:spectrum}, we consider constant Rabi couplings and restrict ourselves to a regime in which the explicitly oscillating dissipative terms average out (see Appendix~\ref{sec:appendG}). In this limit, the resulting rotating-frame generator is time independent.

In the following sections, we use the driven three-level system coupled to a thermal bath as a testbed for the unified master-equation framework presented in Sec.~\ref{sec:thesystem}. For the explicit calculations presented below, we specialize to a Lorentzian spectral density given by
\begin{align}
    J_{\eta,\upsilon}(\omega) = \dfrac{\Gamma_{\eta,\upsilon}}{\pi}\left(\dfrac{\left(\kappa_{\eta,\upsilon}/2\right)^2}{\left(\kappa_{\eta,\upsilon}/2\right)^2 + \left(\omega - \abs{\omega^0_{\eta,\upsilon}}\right)^2}\right)\,.
    \label{eqn:diagspectraldensity}
\end{align}
Here, $\Gamma_{\eta,\upsilon}$ is the on-resonance dissipation rate,  $\kappa_{\eta,\upsilon}$ is the spectral width, and the Lorentzian is centered at the frequency $\abs*{\omega^0_{\eta,\upsilon}}$ of the transition, where $\omega^0_{\eta,\upsilon} = \omega^0_{\eta}-\omega^0_{\upsilon}$. For a bath with nondegenerate mode frequencies, the cross spectral-density function is then given by
\begin{align}
    J_{\eta,\upsilon,\,\eta',\upsilon'}\left(\omega\right) =  e^{i\left(\varphi_{\eta,\upsilon}-\varphi_{\eta',\upsilon'}\right)}\sqrt{J_{\eta,\upsilon}\left(\omega\right)J_{\eta',\upsilon'}\left(\omega\right)}\,.
     \label{eqn:crossspectraldensity}
\end{align}

Section~\ref{sec:protocols} introduces three driving protocols that probe different forms and timescales of the external drive and compares the population dynamics and coherence trajectories for the three different master-equation formulations. Section~\ref{sec:spectrum} presents fluorescence spectra and delineates the parameter regions in which the results for the LF and generalized RF descriptions depart from those for the IME description.

\begin{table*}
\centering
\caption{Parameters for the three driving protocols used in Fig.~\ref{fig:protocol_dynamics}. $\Gamma_0$ is a reference rate that serves as our unit. Frequencies and rates ($\Omega_{\eta,3}^{(0)}$, $\Delta$, $\kappa_{\eta,\upsilon}$, $\Gamma_{\eta,\upsilon}$, and $\omega^L$) are given in units of $\Gamma_0$, the ramp time $t_r$ in units of $\Gamma_0^{-1}$, and the chirp rate $\chi$ in units of $\Gamma_0^2$. The modulation depth $a$ is dimensionless. The parameters common to all three protocols are: the bare transition frequencies $(\omega_{1,3}^{0},\,\omega_{2,3}^{0})=(3,\,2.99)\times10^5 \Gamma_0$; the coherent drive phases $(\phi_{1,3},\,\phi_{2,3})=(0.3,\,-0.2)$; the decay rate $\Gamma_{1,2}=0$; and the coupling phases $\varphi_{1,1} = \pi$ and $\varphi_{2,2}=\varphi_{3,3}=\varphi_{1,3}=\varphi_{2,3}=0$. The drive frequencies satisfy $\omega_{\eta,3}^{d}=\omega_{\eta,3}^{0}-\Delta$, with equal detunings $\Delta_1=\Delta_2=\Delta$.}
\label{tab:protocol_parameters}
\renewcommand{\arraystretch}{1.9}
\setlength{\tabcolsep}{9pt}
\begin{tabular}{ccccc}
\hline\hline
\multirow{2}{*}{Parameter} & \multirow{2}{*}{Units} & Protocol 1 & Protocol 2 & Protocol 3 \\
&  & \footnotesize \shortstack{\textit{Constant}\\\textit{Rabi coupling}} & \footnotesize \shortstack{\textit{Periodically modulated}\\\textit{Rabi coupling}} & \footnotesize \shortstack{\textit{Nonperiodic}\\\textit{(ramped \& chirped)}} \\
\hline
$\Omega_{1,3}^{(0)}$ & $\Gamma_0$ & $2$ & $3.5$ & $4$ \\
$\Omega_{2,3}^{(0)}$ & $\Gamma_0$ & $3.1225$ & $1.5$ & $2$ \\
$\Delta$ & $\Gamma_0$ & $1.5$ & $0.8$ & $-1$ \\
$\kappa_{1,3}=\kappa_{2,3}$ & $\Gamma_0$ & $15$ & $20$ & $25$ \\
$\kappa_{1,1}=\kappa_{2,2}=\kappa_{3,3}$ & $\Gamma_0$ & $20$ & $30$ & $30$ \\
$(\Gamma_{1,3},\Gamma_{2,3})$ & $\Gamma_0$ & $(0.8,0.4)$ & $(0.45,0.45)$ & $(0.5,0.5)$ \\
$(\Gamma_{1,1},\Gamma_{2,2},\Gamma_{3,3})$ & $\Gamma_0$ & $(0.15,0.30,0.15)$ & $(0.20,0.10,0.10)$ & $(0.25,0.125,0.125)$ \\
$\omega^L$ & $\Gamma_0$ & --- & $0.25$ & $0.10$ \\
$t_r$ & $\Gamma_0^{-1}$ & --- & --- & $2.5$ \\
$a$ & --- & --- & --- & $0.85$ \\
$\chi$ & $\Gamma_0^2$ & --- & --- & $0.015$ \\
\hline\hline
\end{tabular}
\end{table*}

\section{\label{sec:protocols}Dynamics for three distinct\texorpdfstring{\\}{ }driving protocols}

In this section, we use the driven three-level $\Lambda$ system introduced in Sec.~\ref{sec:threelevel} to examine the dynamics obtained from the IME, LF, and generalized RF master equations under three driving protocols of increasing temporal complexity: constant-Rabi-coupling (continuous-wave), periodically-modulated-Rabi-coupling, and nonperiodic (ramped and chirped) driving. Throughout this section, the analysis is performed in the rotating frame defined in Sec.~\ref{sec:threelevel} for all three master-equation formulations. The frame in which the results are presented should not be confused with the ``labeling'' of the master equation (LF versus RF versus IME).

\subsection{Definition of the driving protocols}
Since the transformation to the rotating frame and the application of the rotating-wave approximation remove the rapid oscillations at frequencies $\omega_{\eta,3}^{d}$, the only time dependence enters through $\Omega_{1,3}(t)$ and $\Omega_{2,3}(t)$. The three protocols are distinguished by the time dependence of the Rabi couplings $\Omega_{\eta,3}(t)$.

\paragraph*{Protocol 1: Constant-Rabi-coupling driving.} 
Protocol 1 uses $\Omega_{\eta,3}(t)=\Omega_{\eta,3}^{(0)}$. In this continuous-wave limit, the rotating-frame Hamiltonian $H_s^R$ is time independent. We thus choose a dynamical invariant that is time independent and commutes with $H_s^R$. Since the resulting dynamics are governed by a time-independent Hamiltonian, the generalized RF master equation [Eqs.~\eqref{eqn:unifiedgenME} and \eqref{eqn:disscoefficient-RF}] reduces to the conventional RF description (see the discussion in the second-to-last paragraph of Sec.~\ref{sec:thesystem}). Protocol 1 provides a reference for the Rabi-coupling-modulated Protocols 2 and 3. Protocol 1 is also used for the fluorescence spectrum analysis presented in Sec.~\ref{sec:spectrum}.

\paragraph*{Protocol 2: Periodically-modulated-Rabi-coupling driving.} 
Protocol 2 is characterized by a periodic modulation of the $(3{\leftrightarrow}1)$ Rabi coupling while holding the $(3{\leftrightarrow}2)$ Rabi coupling constant,
\begin{align}
    \Omega_{1,3}(t)
    =\Omega_{1,3}^{(0)}\sin^2\!\left(\omega^L t\right),
    \qquad
    \Omega_{2,3}(t)=\Omega_{2,3}^{(0)},
    \label{eqn:slowperiodicdrive}
\end{align}
with envelope period $T_L=\pi/\omega^L$. We assume $\omega^L\ll\omega_{\eta,3}^{d}$ so that the drive in the laboratory frame exhibits three timescales that are set by the rapid drive frequencies $\omega_{1,3}^{d}$ and $\omega_{2,3}^{d}$ and the slower envelope modulation frequency $\omega^L$. Unlike in the constant-Rabi-coupling case, $H_s^R(t)$ is explicitly time dependent, implying that the complex transition amplitudes $\xi_{m,n,\eta,\upsilon}(t)$ and the drive-modified frequencies $\omega_{m,n,\eta,\upsilon}(t)$ evolve during the protocol.

\paragraph*{Protocol 3: Nonperiodic driving.} 
Protocol 3 uses $\Omega_{2,3}(t)=\Omega_{2,3}^{(0)}$ and applies a smooth ramp with a chirped Rabi coupling modulation to the $(3{\leftrightarrow}1)$ transition,
\begin{align}
    \Omega_{1,3}(t)
    =\Omega_{1,3}^{(0)}
    \left(1-e^{-(t/t_r)^2}\right)
    \left[1-a\sin^2\left(\omega^L t+\chi t^2\right)\right],
    \label{eqn:nonperiodicdrive}
\end{align}
where $t_r>0$ sets the ramp turn-on time, $0<a<1$ is the modulation depth, and $\chi\ne0$ is the chirp rate. For a positive chirp ($\chi>0$), the instantaneous angular frequency $2(\omega^L+2\chi t)$ of the modulation increases linearly with time, producing progressively shorter modulation time periods and a nonperiodic rotating-frame Hamiltonian $H_s^R(t)$.

Protocol 3 illustrates the applicability of the IME beyond stationary and periodic drives. We reiterate that the dissipative parts of the LF master equation completely neglect the time dependence of the driving protocol. Importantly, while the conventional RF master equation cannot be applied to Protocol 3 because $H_s^R(t)$ is nonperiodic in time, the generalized RF master equation introduced in this work can be applied.

\subsection{Dynamics of the density-matrix elements}
For all three protocols, we initialize the system in $\rho(0)=\ket{1}\bra{1}$ at zero bath temperature. All frequencies and rates are expressed in units of a reference rate $\Gamma_0$, with time in the corresponding inverse unit $\Gamma_0^{-1}$. For context, we note that $\Gamma_0 = 1\,\text{GHz}$ is a typical scale for excited-state decay rates in optical emitters, including semiconductor quantum dots~\cite{Lodahl2015}, certain color centers~\cite{Rogers2014}, and fluorescent molecules~\cite{Toninelli2021}. The protocol-specific parameters are listed in Table~\ref{tab:protocol_parameters}.

\begin{figure}
\centering
\includegraphics[width=\columnwidth]{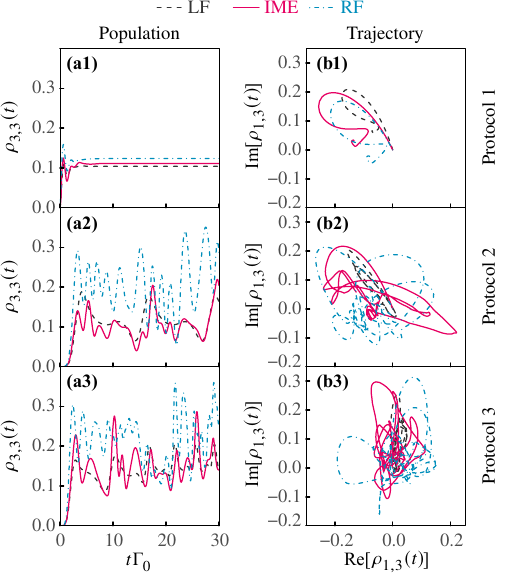}
\caption{Population dynamics and coherence trajectories for Protocol 1 (constant-Rabi-coupling driving, top row), Protocol 2 (periodically-modulated-Rabi-coupling driving, middle row), and Protocol 3 (nonperiodic driving, bottom row). The left and right columns show the excited-state population $\rho_{3,3}(t)$ and the coherence trajectories $\rho_{1,3}(t)$ in the complex plane, respectively. The dynamics are obtained using the LF (dashed lines), IME (solid lines), and generalized RF (dash-dotted lines) master equations; the latter reduces to the conventional RF equation for Protocol 1. The parameters are given in Table~\ref{tab:protocol_parameters}.}
\label{fig:protocol_dynamics}
\end{figure}

This subsection examines the population dynamics and coherence trajectories for the three driving protocols introduced in the previous subsection. Figure~\ref{fig:protocol_dynamics} shows the excited-state population $\rho_{3,3}(t)$ as a function of the dimensionless time $\Gamma_0t$ in the left column and the complex-plane trajectory of the coherence $\rho_{1,3}(t)$ in the right column. The trajectory axes show $\Re[\rho_{1,3}(t)]$ and $\Im[\rho_{1,3}(t)]$. The top, middle, and bottom rows correspond to Protocols~1, 2, and~3, respectively. The LF, IME, and generalized RF results are represented by dashed, solid, and dash-dotted lines, respectively. The excited-state population $\rho_{3,3}(t)$ illustrates the competition between coherent driving and dissipation, while the complex-plane trajectory of $\rho_{1,3}(t)$ captures how the different dissipation coefficients affect both the magnitude and the phase of the coherence. The examples presented in this subsection are chosen to highlight differences between the three master-equation frameworks for parameter combinations for which positivity of the density matrix is a given. We carefully benchmarked our formulations and codes by checking that the three master-equation formulations yield essentially identical results for appropriately chosen parameter combinations.

For the constant-Rabi-coupling drive (Protocol 1),  Fig.~\ref{fig:protocol_dynamics}(a1) shows that the population $\rho_{3,3}(t)$ initially increases and then approaches a stationary value for all three master-equation approaches. While the populations reach their steady-state values at approximately the same time for all three master-equation approaches, their values differ by a few percent: the stationary LF and generalized RF populations are approximately $6.5\%$ below and $11.2\%$ above the IME population, respectively. Differences can also be seen in the coherence trajectories [Fig.~\ref{fig:protocol_dynamics}(b1)]: the IME trajectory extends to more negative $\Re[\rho_{1,3}(t)]$ than the LF and generalized RF trajectories. The generalized RF trajectory acquires negative $\Im[\rho_{1,3}(t)]$ and converges to a stationary coherence whose imaginary part has the opposite sign to that of $\Im[\rho_{1,3}(t)]$ for the IME and LF approaches. Overall, the dynamics of $\rho_{3,3}(t)$ and $\rho_{1,3}(t)$ obtained by the three master-equation approaches may be described as being qualitatively similar, with quantitative differences existing in both the transient and steady-state regimes.

To quantify deviations from the IME approach, we compute the relative root-mean-square (RMS) deviation
\begin{align}
\epsilon_X[q]
=
\left[
\frac{\displaystyle\int_0^T
\lvert q_X(t)-q_{\mathrm{IME}}(t)\rvert^2\,dt}
{\displaystyle\int_0^T
\lvert q_{\mathrm{IME}}(t)\rvert^2\,dt}
\right]^{1/2}\,,
\label{eqn:rms}
\end{align}
where $X\in\{\mathrm{LF},\mathrm{RF}\}$ labels the approach and $q_X(t)$ denotes either the corresponding population $\rho_{3,3}(t)$ or the complex coherence $\rho_{1,3}(t)$, and $T=30\Gamma_0^{-1}$. For the periodically modulated and nonperiodic drives (Protocols~2 and~3), we can identify the following common characteristics. (i) The population $\rho_{3,3}(t)$ obtained by the LF approach follows the population obtained by the IME approach quite closely, with the LF approach ``smoothing'' the oscillations resolved by the IME approach, with $\epsilon_{\mathrm{LF}}[\rho_{3,3}]=26\%$ and $21\%$ for Protocols 2 and 3, respectively. This behavior is consistent with the LF approach neglecting drive-induced evolution over the bath correlation time and restricting bath sampling to bare transition frequencies. (ii) The population $\rho_{3,3}(t)$ obtained by the generalized RF approach, in contrast, oscillates at essentially the same frequencies as the population obtained by the IME approach; however, the generalized RF oscillation amplitudes are notably larger than the IME oscillation amplitudes [see Figs.~\ref{fig:protocol_dynamics}(a2) and~(a3)]. Although the generalized RF treatment retains drive-modified bath sampling, it discards nonresonant couplings between distinct dynamical transitions. The larger generalized RF oscillation amplitudes, with $\epsilon_{\mathrm{RF}}[\rho_{3,3}]=111\%$ for Protocol 2 and $61\%$ for Protocol 3, suggest that these couplings suppress population modulation in the IME for the parameters considered. (iii) The LF coherence trajectories are confined to a smaller region of the complex plane than the IME and generalized RF trajectories [Figs.~\ref{fig:protocol_dynamics}(b2) and~(b3)]. We find $\epsilon_{\mathrm{LF}}[\rho_{1,3}]=61\%$ for Protocol 2 and $55\%$ for Protocol 3. By modifying the dissipative weights of the invariant-basis transitions, the LF approach alters the balance between coherence generation and decay, reshaping both the magnitude and phase of the bare-basis coherence compared to the IME.

The three protocols progress from a time-independent drive to a time-periodic drive to a nonperiodic drive, demonstrating how the treatment of the dissipation coefficients alters both the population dynamics and the coherence trajectories. In the cases where differences arise between the different treatments, the IME provides the most accurate description of the transient and steady-state features, provided the density matrix is positive. In Sec.~\ref{sec:spectrum}, we return to the constant-Rabi-coupling driving protocol and use the steady-state fluorescence spectrum to compare the IME description with the LF and generalized RF descriptions for a wide range of parameters.

\section{\label{sec:spectrum}Analysis of fluorescence spectra}
Fluorescence spectroscopy provides an experimentally accessible probe of driven quantum dynamics, with frequencies, linewidths, and asymmetries being characteristics of the spectra~\cite{grove1977,ates2009,Mollow1969,Glauber1963}. Here, we compare selected transition-resolved spectra obtained within the IME, LF, and generalized RF approaches to reveal dynamical differences that steady-state populations alone may conceal. In this section, we omit bath-induced Lamb shifts for simplicity. We expect that inclusion of the Lamb shifts would introduce small quantitative shifts in the fluorescence spectra but would not affect their qualitative behavior. Throughout this section, we use Protocol 1, the initial state $\ket{1}\bra{1}$, and the bath temperature $T_b = 0$.

\subsection{Definition}
For the transition operator $\sigma_{\eta,\upsilon}$, we define the incoherent part of the correlation function in the rotating frame through
\begin{align}
    g^{\mathrm{inc}}_{\eta,\upsilon}
    (\tau)=\lim_{t_0\rightarrow\infty}\expval{\delta\sigma_{\eta,\upsilon}^{\Dag}(t_0)\delta\sigma_{\eta,\upsilon}(t_0+\tau)}\,,
    \label{eqn:connected_correlation}
\end{align}
where $\delta\sigma_{\eta,\upsilon}(t) =\sigma_{\eta,\upsilon}(t) -\expval{\sigma_{\eta,\upsilon}}_{\mathrm{ss}}$ and $\expval{\sigma_{\eta,\upsilon}}_{\mathrm{ss}}$ denotes the steady-state expectation value. The subtraction of the steady-state mean removes the elastic contribution, giving the incoherent spectrum. Specifically, the incoherent spectrum is given by\vspace{-1em}
\begin{align}
   S^{\mathrm{inc}}_{\eta,\upsilon}
   (\omega)=\frac{1}{2\pi}\int_{-\infty}^{\infty}d\tau\,e^{-i\omega\tau}g^{\mathrm{inc}}_{\eta,\upsilon}
   (\tau)\,.
    \label{eqn:incoherent_spectrum}
\end{align}

We evaluate the correlation function using the quantum regression theorem~\cite{lax1963,steck_quantum_atom_optics} and the generator $\mathcal L$ that governs the density-matrix evolution, i.e., $\dot{\rho}(t) = \mathcal L [\rho(t)]$. Assuming $\mathcal L$ is time independent (see Appendix~\ref{sec:appendG} for the approximations needed to realize a time-independent $\mathcal L$), we introduce the operator $\Lambda_{\eta,\upsilon}(\tau)$, which satisfies~\cite{basak2026}
\begin{align}
    \frac{d}{d\tau}\Lambda_{\eta,\upsilon}(\tau)
    &=
 \mathcal L\!\left[\Lambda_{\eta,\upsilon}(\tau)\right]
 \label{eqn:regression_correlation-a}
\intertext{and}
    \Lambda_{\eta,\upsilon}(0)
    &=
    \rho_{\mathrm{ss}}\,
    \delta\sigma_{\eta,\upsilon}^{\Dag}\,,
    \label{eqn:regression_correlation-b}
\intertext{so that}
    g^{\mathrm{inc}}_{\eta,\upsilon}(\tau)
    &=
    \Tr\left[
        \delta\sigma_{\eta,\upsilon}
        \Lambda_{\eta,\upsilon}(\tau)
    \right]\,.
    \label{eqn:regression_correlation}
\end{align}
Here, $\rho_{\mathrm{ss}}$ is the steady-state density matrix.

We assume that the time-independent generator $\mathcal{L}$ is diagonalizable (admits a spectral decomposition), with a simple stationary eigenvalue $\lambda_0=0$ and $\Re[\lambda_\ell]<0$ for $\ell\neq0$. The right eigenmatrix that corresponds to the stationary eigenvalue represents the stationary mode. The other eigenmatrices represent decaying dynamical modes. For a system that possesses a unique stationary mode (or steady state) and decaying dynamical (or transient) modes, we can write the correlation function and the resulting spectrum as~\cite{basak2026_analyticalfluorescence,scarlatella2019}
\begin{align}
g^{\mathrm{inc}}_{\eta,\upsilon}(\tau) &= \sum_{\ell\ne0}C_\ell^{\eta,\upsilon}e^{\lambda_\ell\tau}
\label{eqn:correlation_modes} 
\intertext{and\vspace{-1em}}
S^{\mathrm{inc}}_{\eta,\upsilon}(\omega) &= \frac{1}{\pi}\Re\left[ \sum_{\ell\ne0} \frac{C_\ell^{\eta,\upsilon}}{i\omega-\lambda_\ell}\right]\,, 
\label{eqn:spectrum_modes} 
\end{align} 
where, with $R_\ell$ and $L_\ell$ representing the right and left eigenmatrices of $\mathcal L$, the spectral weights are given by~\cite{basak2026_analyticalfluorescence,scarlatella2019}
\begin{align}
    C_\ell^{\eta,\upsilon}
    &= 
    \Tr(\delta\sigma_{\eta,\upsilon}R_\ell)
    \Tr
    (L_\ell^{\Dag}\rho_{\rm ss}\delta\sigma_{\eta,\upsilon}^{\Dag}).
    \label{eqn:spectralweights}
\end{align}
Writing the eigenvalues associated with the dynamical modes as $\lambda_\ell=-\gamma_\ell+i\upsilon_\ell$,  Eq.~\eqref{eqn:spectrum_modes} decomposes the fluorescence spectrum into a sum of contributions for each mode, which manifest as distinct peaks in the spectrum. The quantity $\gamma_\ell$ sets the relaxation rate of the $\ell$th dynamical mode and determines the corresponding spectral linewidth. For weak dissipation compared to the drive strength, $\gamma_\ell$ is to leading (first) order governed by the dissipative part of the generator. The quantity $\upsilon_\ell$ sets the oscillation frequency of the mode and fixes the resonance position of the spectral peak. For weak dissipation compared to the drive strength, $\upsilon_\ell$ is determined to leading (zeroth) order by the energy differences between the dressed states, and thus by the generalized Rabi couplings.

The generally complex coefficients $C_\ell^{\eta,\upsilon}$ [see Eq.~\eqref{eqn:spectralweights}] contain information about the overlap of the dynamical modes $\smash{R_\ell}$ and $\smash{L_\ell^{\Dag}}$ with both the stationary state $\rho_{\mathrm{ss}}$ and the transition operators $\smash{\delta\sigma_{\eta,\upsilon}}$ and $\smash{\delta\sigma_{\eta,\upsilon}^{\Dag}}$ associated with the transition that is being probed in the spectrum~\cite{basak2026_analyticalfluorescence,scarlatella2019}. The real part of $C_\ell^{\eta,\upsilon}$ determines the weight of the purely Lorentzian peak, while the imaginary part introduces a dispersive character, thereby governing the peak's asymmetry. In the generalized RF approach, due to the secular approximation, the $C_\ell^{\eta,\upsilon}$ are strictly real; consequently, the peaks are always purely Lorentzian with no dispersive contribution~\cite{basak2026}. Since changing the master equation can alter both the eigenvalues $\lambda_\ell$ and the spectral weights $C_\ell^{\eta,\upsilon}$, different dissipation coefficients---as encountered in the LF, generalized RF, and IME approaches---can produce differences in spectral peak positions, widths, relative weights, and asymmetries.

In general, the spectral response is obtained from a two-time correlation function that depends on both $t$ and a reference time $t_0$ (or, in the frequency domain, two frequencies $\omega$ and $\omega_0$). Here, we focus on the stationary limit $t_0 \to \infty$, in which the system density matrix has reached its steady state and the spectrum depends only on the relative delay $\tau = t - t_0$, reducing to a single-frequency function $S(\omega)$. We therefore restrict the analysis to a time-independent generator. For this purpose, we choose a constant-Rabi-coupling periodic drive, which makes the Hamiltonian and the chosen dynamical invariant in the rotating frame time independent. To eliminate the remaining explicit time dependence in the dissipative terms, we make three approximations (see Appendix~\ref{sec:appendG} for details). (i) We set the system--bath coupling $g_{1,2,k}$ to zero to eliminate the time dependence that is generated by nonresonant couplings, which involve the $(1\leftrightarrow2)$ channel. Correspondingly, this decay/excitation channel and all associated cross-dissipative terms are absent. (ii) We neglect rapidly oscillating cross-dissipative terms. For the system under consideration, these terms oscillate at frequencies that are well separated from the relevant dissipative scales and thus average to zero to a very good approximation over the timescale of interest. (iii) Although the drive frequencies $\smash{\omega^d_{1,3}}$ and $\smash{\omega^d_{2,3}}$ associated with $\Omega_{1,3}(t)$ and $\Omega_{2,3}(t)$ (as well as the corresponding bare transition frequencies) are much larger than the relevant dissipative rates and drive strengths, their difference $\smash{\abs*{\omega^d_{1,3}-\omega^d_{2,3}}}$ does not need to be. Consequently, cross-dissipative terms that couple the $(1{\leftrightarrow}3)$ and $(2{\leftrightarrow}3)$ transitions, which oscillate at the frequency $(\omega^d_{1,3}-\omega^d_{2,3})$, can survive the standard secular approximation and need to be considered separately. To neglect these remaining cross terms, we impose the additional separation-of-scales condition $|{\omega^d_{1,3}-\omega^d_{2,3}}|\gg\sqrt{\Gamma_{1,3}\Gamma_{2,3}}$.

\subsection{\label{subsec:specsign}Spectral signatures}
\begin{figure*}
\centering
\vspace{1.5cm}
\includegraphics{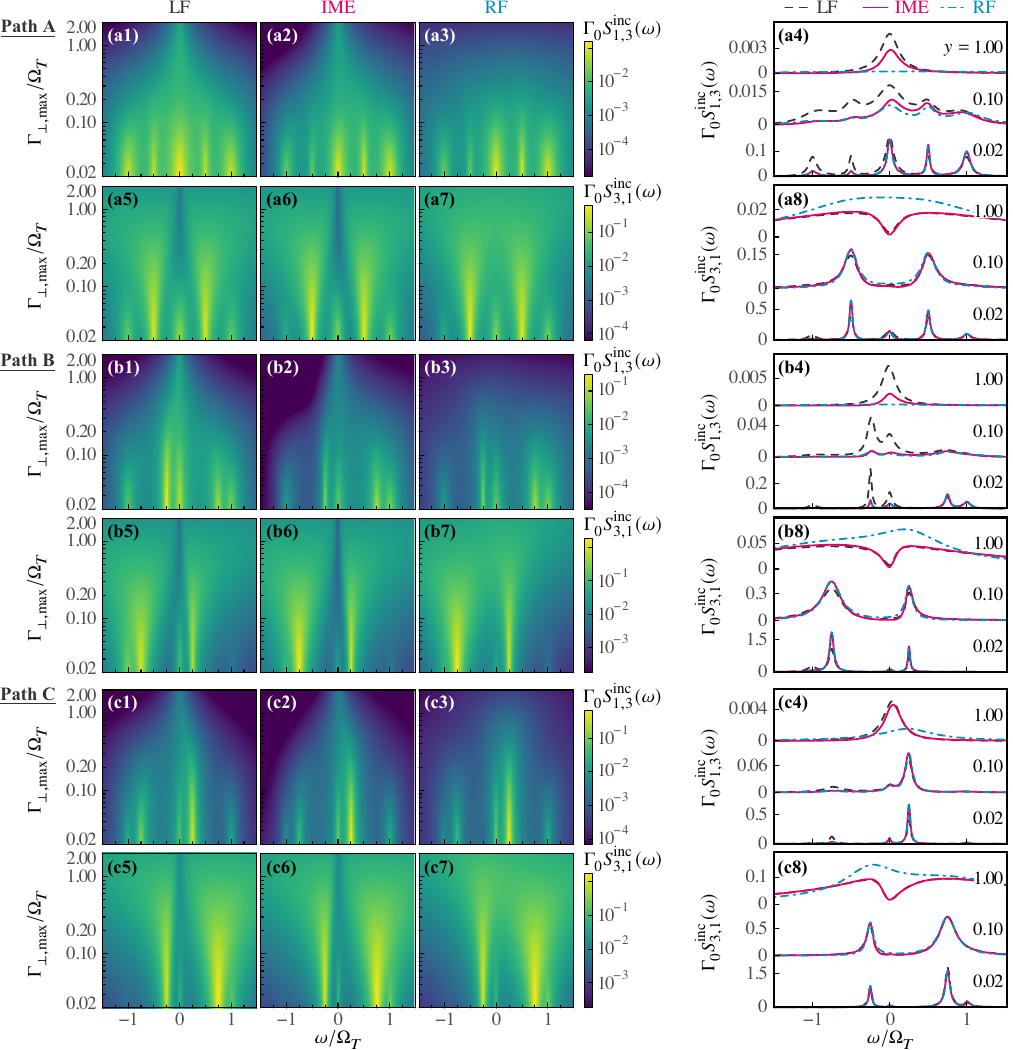}
\caption{Incoherent emission and absorption spectra for Paths A [(a1)--(a8)], B [(b1)--(b8)], and C [(c1)--(c8)].
For each path, the panels labeled 1--4 show the emission spectrum $S^{\mathrm{inc}}_{1,3}(\omega)$ while the panels labeled 5--8 show the absorption spectrum $S^{\mathrm{inc}}_{3,1}(\omega)$.
The first, second, and third columns correspond to results from the LF, IME, and generalized RF approaches, respectively; the fourth column shows cuts at $y=\Gamma_{\perp,\max}/\Omega_T=0.02$, $0.1$, and $1$ using a linear intensity scale.
It can be seen that the IME spectrum is captured reasonably well by the LF approach for large $y$ and by the generalized RF approach for small $y$.
For all three paths, $\Omega_T/\kappa_{\min}\approx0.0889$ and $\Gamma_{\parallel,\max}/\Omega_T=0.01$.
The path constraints and bath parameters are given in Table~\ref{tab:paths}, and the bath temperature is $T_b=0$.\vspace{1.5cm}}
\label{fig:transition}
\end{figure*}

\begin{table}
\centering
\caption{Parameter constraints that define Paths A, B, and C. The columns specify the drive ratio, detuning-to-generalized-Rabi-coupling ratio, and relative transverse and longitudinal rates. $\Gamma_{\perp,\max}$ is defined as $\max{(\Gamma_{1,3},\Gamma_{2,3})}$ (recall, $\Gamma_{1,2}=0$ throughout Sec.~\ref{sec:spectrum}) and $\Gamma_{\parallel,\max}$ as $\max{(\Gamma_{1,1},\Gamma_{2,2},\Gamma_{3,3})}$. For all paths, $\kappa_{1,3}=\kappa_{2,3}=20\Gamma_0$ and $\kappa_{1,1}=\kappa_{2,2}=\kappa_{3,3}=30\Gamma_0$. $\kappa_{\min}$ is defined as $\kappa_{\min}=\min{(\kappa_{1,3},\kappa_{2,3},\kappa_{1,1},\kappa_{2,2},\kappa_{3,3})}$. The diagonal coupling phases are $\varphi_{1,1}=\pi$ and $\varphi_{2,2}=\varphi_{3,3}=0$.}
\label{tab:paths}

\setlength{\tabcolsep}{8pt}

\begin{tabular}{ccccc}
\toprule
Path
& $\dfrac{\Omega_{2,3}}{\Omega_{1,3}}$
& $\dfrac{\Delta}{\Omega_T}$
& $\dfrac{(\Gamma_{1,3},\Gamma_{2,3})}
          {\Gamma_{\perp,\max}}$
& $\dfrac{(\Gamma_{1,1},\Gamma_{2,2},\Gamma_{3,3})}
          {\Gamma_{\parallel,\max}}$
\\
\midrule
A & $1/2$ & $0$
  & $\tupleTwo{1}{1}$
  & $\tupleThree{1}{1/2}{1/2}$
\\
B & $1$ & $-1/2$
  & $\tupleTwo{1}{1}$
  & $\tupleThree{1/4}{1}{1}$
\\
C & $2$ & $+1/2$
  & $\tupleTwo{1}{1/2}$
  & $\tupleThree{1/2}{1}{1/2}$
\\
\bottomrule
\end{tabular}
\end{table}

This section determines how the competition between coherent driving and dissipation manifests in the transition-resolved spectra. We define $\Gamma_{\perp,\max}=\max{(\Gamma_{1,3},\Gamma_{2,3})}$,  $\Gamma_{\parallel,\max}=\max{(\Gamma_{1,1},\Gamma_{2,2},\Gamma_{3,3})}$, and $\kappa_{\min}=\min{(\kappa_{1,3},\kappa_{2,3},\kappa_{1,1},\kappa_{2,2},\kappa_{3,3})}$. The $(1\leftrightarrow2)$ bath channel is absent throughout this section since we set $g_{1,2,k}=0$ for all bath modes $k$ to make the generator time independent. The full parameter space includes the Rabi couplings $\Omega_{1,3}$ and $\Omega_{2,3}$, the common detuning $\Delta_1=\Delta_2=\Delta$, the transverse decay rates $\Gamma_{1,3}$ and $\Gamma_{2,3}$, the longitudinal coupling strengths $\Gamma_{\eta,\eta}$ ($\eta=1,2,3$), and the bath spectral widths $\kappa_{\eta,\upsilon}$. We consider three parameter paths, Paths A, B, and C (Table~\ref{tab:paths}). Path A fixes $\Omega_{2,3}=\Omega_{1,3}/2$, $\Delta=0$, $\Gamma_{2,3}=\Gamma_{1,3}$, and $\Gamma_{2,2}=\Gamma_{3,3}=\Gamma_{1,1}/2$. Paths B and C are characterized by $\Delta/\Omega_T=-1/2$ and $+1/2$, respectively (recall, $\Omega_T$ denotes the generalized Rabi coupling), and use different drive ratios and relative transverse and longitudinal coupling strengths to assess how the spectra depend on these ratios. For all three paths, $\Omega_T/\kappa_{\min}\approx0.0889$ and $\Gamma_{\parallel,\max}/\Omega_T=0.01$.

Figure~\ref{fig:transition} shows the incoherent emission spectrum $S^{\rm inc}_{1,3}(\omega)$ and the incoherent absorption spectrum $S^{\rm inc}_{3,1}(\omega)$ for Path A [(a1)--(a8)], Path B [(b1)--(b8)], and Path C [(c1)--(c8)]. For each path, the first and second rows show the emission and absorption spectra, respectively. The first three columns present the LF, IME, and generalized RF results as functions of the normalized spectral frequency $\omega/\Omega_T$ (linear scale) and the transverse decay-to-drive ratio $y=\Gamma_{\perp,\max}/\Omega_T$ (logarithmic scale). The spectral intensities span more than four orders of magnitude and are represented using logarithmic color scales. 

For all three parameter paths, the IME spectra show a crossover from a drive-dominated regime with five resolved peaks at small $y$ to a dissipation-dominated regime characterized by a single central feature---a peak in the emission spectrum $S^{\mathrm{inc}}_{1,3}(\omega)$ and a dip in the absorption spectrum $S^{\mathrm{inc}}_{3,1}(\omega)$--- around $\omega = 0$ at large $y$. At small $y$ ($y \ll 1$), where the coherent drive dominates over dissipation, the five peaks correspond to distinct dressed-state transitions, representing a three-level generalization of the two-level Mollow triplet~\cite{Mollow1969, basak2026}. Physically, although the fluorescence spectrum probes a single bare atomic transition, i.e., emission from $\ket{3}\to\ket{1}$ via $\sigma_{1,3}$ or absorption from $\ket{1}\to\ket{3}$ via $\sigma_{3,1}$, the strong coherent fields (comparatively large $\Omega_T$) dress the bare atomic levels, yielding three hybridized eigenstates (dressed states) whose energy splittings are set by the generalized Rabi coupling $\Omega_T$ and the detuning $\Delta$. Because each bare state can be written as a superposition of these dressed states, a single bare transition corresponds to transitions between all pairs of contributing dressed states. When the energy scales associated with the dissipation are much smaller than the dressed-state energy splittings ($y \ll 1$), the dressed-state transitions are weakly damped and mostly uncoupled from one another. This implies that each transition between dressed states contributes a distinct, well-resolved peak centered at the corresponding dressed-state energy difference.

For the resonant Path A ($\Delta_1=\Delta_2=\Delta=0$), e.g., the rotating-frame Hamiltonian possesses one dark state at energy zero (which has no admixture of the  excited state $\ket{3}$) and two bright dressed states at energies $\pm\Omega_T/2$. The differences of these three dressed-state energies yield five resonance frequencies at $\omega/\Omega_T = 0$, $\pm 1/2$ (bright--dark transitions), and $\pm 1$ (bright--bright transitions). As $y$ increases, enhanced dissipation broadens the spectral peaks until their widths are comparable to the dressed-state splittings, washing out the sidebands (i.e., the peaks at $\omega \ne 0$). At decay-to-drive ratios $y \gtrsim 1$, the dissipative rates are comparable to or exceed the dressed-state energy splittings ($\Gamma_{\perp,\max} \gtrsim \Omega_T$), implying that the system decays faster than the coherent drive can complete a Rabi oscillation. Consequently, dissipation combines the dressed states into collective modes, implying that the dynamics can no longer be described in terms of isolated transitions between dressed states. As dissipation suppresses the oscillations responsible for the sidebands, the distinct side peaks disappear. The central feature at $\omega = 0$ reflects the remaining relaxation dynamics: a comparatively slowly decaying contribution produces the relatively narrow central feature while the rapidly decaying contributions produce the broad background.

Comparison of the emission and absorption spectra in Fig.~\ref{fig:transition} shows that $S^{\mathrm{inc}}_{1,3}(\omega)$ and $S^{\mathrm{inc}}_{3,1}(\omega)$ behave, in some sense, as ``inverses.'' Because emission (due to the operator $\sigma_{1,3}$) requires initial population in the excited state $\ket{3}$ whereas absorption (due to the operator $\sigma_{3,1}$) probes transitions that start from the ground state $\ket{1}$, the two operators project differently onto the dark and bright states. At small $y$, the bright--dark sidebands ($\omega/\Omega_T = \pm 1/2$ for Path A) are suppressed in the emission spectrum but dominate in the absorption spectrum. At large $y$, in contrast, the IME emission spectrum shows a pronounced central maximum at $\omega = 0$ whereas the absorption spectrum exhibits a sharp central dip within a broad background. Overall, the LF, IME, and generalized RF spectra exhibit similarly broad features. Differences in the intensities and line shapes of the spectra, however, motivate more detailed comparisons.

The fourth (rightmost) column of Fig.~\ref{fig:transition} shows the spectral intensity on a linear scale for selected cuts at $y=0.02$, $0.1$, and $1$. At small $y=0.02$, the generalized RF and IME spectra agree well. The LF spectra, in contrast, deviate from the IME spectra, distorting the relative intensities and asymmetry between the positive- and negative-frequency emission sidebands (the LF approach also overestimates the overall emission intensity as is most visible for Path B). At the intermediate value of $y=0.1$, the peaks are broader than at $y=0.02$ and begin to overlap. The generalized RF approach underestimates the central emission intensity and overestimates the central absorption intensity for all three paths. The LF and IME emission spectra show, as at $y=0.02$, discrepancies, most visibly for Path B. At $y=1$, the LF approach captures the central emission maximum and absorption dip seen in the IME qualitatively. However, the LF approach overestimates the central emission intensity along Paths A and B and agrees more closely with the IME along Path C. The LF  absorption spectra are close to the IME absorption spectra for all three paths. The generalized RF approach misses the pronounced central emission maximum and absorption dip, predicting strongly suppressed emission for Paths A and B, a lower and broader emission feature for Path C, and a broad absorption maximum without a central dip.

In what follows, we aim to gain additional insights into the spectra by connecting directly to Eqs.~\eqref{eqn:regression_correlation-a}-\eqref{eqn:spectralweights}. The good agreement between the generalized RF and IME spectra at small $y$ for Paths A, B, and C [Figs.~\ref{fig:transition}(a4), (a8), (b4), (b8), (c4), and~(c8)] motivates us to use the generator $\mathcal{L}$ of the generalized RF approach as opposed to the generator of the IME to interpret the spectral features in the strong-driving regime ($y\ll 1$). The rationale behind our analysis is that the generator of the generalized RF approach has a simpler structure than the generator of the IME approach. The agreement between the IME and generalized RF spectra at small $y$ indicates that the couplings omitted by the secular approximation have a small effect on the spectra in this regime. 

In the dressed-state basis, the generalized RF generator separates the operator space into sectors with distinct dressed-state transition frequencies. The population sector contains the diagonal projection operators $\ket{\mu_m}\bra{\mu_m}$ and is decoupled from the six coherence sectors, which each contain one of the off-diagonal transition operators $\ket{\mu_m}\bra{\mu_n}$ ($m \neq n$). Consequently, the steady state within the generalized RF approach is diagonal, i.e., we have $\rho_{\mathrm{ss}} = \sum_m p_m^{\mathrm{ss}}\ket{\mu_m}\bra{\mu_m}$. It is important to note that the sector separation holds for all $y$ in the generalized RF approach. Moreover, the sector separation of the generalized RF generator  provides a leading-order approximation to the IME generator when dissipation is weak compared to  the separation between distinct transition frequencies ($y \ll 1$). Within the population sector, dissipation mutually couples the dressed-state populations, so that the corresponding right eigenmatrices take the form
\begin{align}
    R_{\ell} = \sum_{m} c_{\ell,m} \ket{\mu_m}\bra{\mu_m}\,,
    \label{eqn:R_population}
\end{align}
where $c_{\ell,m}$ specifies the weight of the diagonal projector $\ket{\mu_m}\bra{\mu_m}$ in $R_{\ell}$. Here, $\ell = 0$ corresponds to the stationary mode $R_0 = \rho_{\mathrm{ss}}$ with $\lambda_0 = 0$. The two decaying population modes ($\ell \neq 0$) have purely real eigenvalues, i.e., $\lambda_{\ell} = -\gamma_{\ell}$ ($\upsilon_{\ell} = 0$), and describe relaxation without oscillation; because $\upsilon_{\ell} = 0$, their spectral contributions are centered at $\omega = 0$ and their widths are set by $\gamma_{\ell}$.

Each of the coherence sectors describes an oscillation at the dressed-state energy difference $\upsilon_{m,n} = E_n - E_m$. The right eigenmatrix is given by
\begin{align}
    R_{m,n} = \ket{\mu_m}\bra{\mu_n}\,, \qquad m \neq n\,,
    \label{eqn:R_coherence}
\end{align}
with $L_{m,n}^{\Dag} = R_{m,n}^{\Dag}$ and eigenvalue $\lambda_{m,n} = -\gamma_{m,n} + i(E_n - E_m)$, where $\gamma_{m,n}$ is the relaxation rate of the dynamical mode.  The energy difference $E_n - E_m$ sets the oscillation frequency. Substituting the eigenmatrices for the six coherence sectors and the diagonal steady state $\rho_{\mathrm{ss}}$ into Eq.~\eqref{eqn:spectralweights}, the spectral weights [Eqs.~\eqref{eqn:correlation_modes}--\eqref{eqn:spectralweights}] reduce to
\begin{align}
    C_{m,n}^{\eta,\upsilon} = p_m^{\mathrm{ss}}\,\abs{\bra{\mu_n}\sigma_{\eta,\upsilon}\ket{\mu_m}}^2\,.
    \label{eqn:coherence_weights}
\end{align}
Each coherence sector's dynamical mode $(m,n)$ thus contributes an oscillating term $C_{m,n}^{\eta,\upsilon} e^{[-\gamma_{m,n} + i(E_n - E_m)]\tau}$ to $g^{\mathrm{inc}}_{\eta,\upsilon}(\tau)$ [Eq.~\eqref{eqn:correlation_modes}] and a purely Lorentzian peak centered at $\omega = E_n - E_m$ to $S^{\mathrm{inc}}_{\eta,\upsilon}(\omega)$ [Eq.~\eqref{eqn:spectrum_modes}]. Equation~\eqref{eqn:coherence_weights} makes the physical content explicit: the weight of each peak is governed by the steady-state population $p_m^{\mathrm{ss}}$ of the initial dressed state and the squared transition amplitude $\abs{\mel{\mu_n}{\sigma_{\eta,\upsilon}}{\mu_m}}^2$ that connects the two dressed states.

As $y$ increases, the transverse contribution to the relaxation rates of the generalized RF approach grows linearly with $y$. Consequently, the spectral peaks broaden and start to
overlap, as seen at $y=0.1$ in Figs.~\ref{fig:transition}(a4), (a8), (b4), (b8), (c4), and~(c8). In addition to broadening the peaks, increasing $y$ modifies the steady-state populations of the dressed states. Because $\Delta_1 = \Delta_2 = \Delta$ for all three paths, the rotating-frame Hamiltonian  supports a dark state $\ket{\mu_0}$, which has vanishing contribution of the excited state $\ket{3}$  (i.e., $\braket{3}{\mu_0}=0$), for all three paths. The transverse decay transfers population from the bright states $\ket{\mu_{\pm}}$ to the dark state $\ket{\mu_0}$, whereas the longitudinal dephasing transfers population back from $\ket{\mu_{0}}$ to $\ket{\mu_{\pm}}$. Since the longitudinal dephasing rate is constant for Paths A, B, and C, increasing $y$ increases the net population transfer from the bright states to the dark state. Within the generalized RF treatment, the spectral weights are proportional to the initial steady-state population and the squared transition amplitude even as $y$ increases. 

We now apply these considerations to the emission and absorption spectra. For the emission spectrum $S^{\mathrm{inc}}_{1,3}(\omega)$, the squared transition amplitudes are zero for the dark state since $\sigma_{1,3}\ket{\mu_0}=0$ and nonzero for the bright states $\ket{\mu_\pm}$. Hence, as $y$ increases and population transfers from the bright states into the dark state, the spectral weights decrease, reducing the peak heights. Together with the broadening, this causes the generalized RF emission spectrum to flatten and have reduced intensity at $y=1$ [Figs.~\ref{fig:transition}(a4), (b4), and~(c4)]. For the absorption spectrum $S^{\mathrm{inc}}_{3,1}(\omega)$, in contrast, the squared transition amplitudes from the dark state $\ket{\mu_0}$ to the bright states $\ket{\mu_\pm}$ are nonzero because $\sigma_{3,1}$ couples directly to the $\ket{1}$ component of $\ket{\mu_0}$. Hence, as $y$ increases, the accumulation of population in the dark state increases the spectral weights of the peaks associated with the dark-to-bright transitions while reducing the spectral weights of the peaks associated with the bright-to-dark and bright-to-bright transitions. Together with the broadening, this causes the peaks to merge constructively into a single broad maximum around $\omega=0$ [Figs.~\ref{fig:transition}(a8), (b8), and~(c8)].

The IME retains the nonsecular couplings between distinct frequency sectors that the generalized RF approach discards. As $y$ increases and the dissipative rates become comparable to the dressed-state frequency separations, the inter-sector couplings can no longer be neglected. In this regime, the dressed state populations and coherences are mixed within the IME approach. Consequently, the spectra can no longer be understood in terms of isolated transitions between pairs of dressed states. Instead, the spectra are governed by the dynamical modes of the IME generator, each of which can be written as a linear combination of both diagonal projection operators $\ket{\mu_m}\bra{\mu_m}$ and off-diagonal transition operators $\ket{\mu_m}\bra{\mu_n}$ ($m \neq n$). At comparatively large decay-to-drive ratios ($y \gtrsim 1$), the generator is dominated by dissipation. In this regime, the nonsecular dissipative couplings hybridize the dressed-state population and coherence sectors into two classes of dynamical modes based on their relaxation rates $\gamma_\ell$: fast-decaying modes with relaxation rates $\gamma_\ell \sim \Gamma_{\perp,\max}$ and slow-decaying modes with relaxation rates $\gamma_\ell < \Gamma_{\perp,\max}$. The slow relaxation is governed by the dephasing rate ($\Gamma_{\parallel,\mathrm{max}}$) together with coherent state transfer that is continuously damped by $\Gamma_{\perp,\max}$ and converted into slow, incoherent relaxation. Because the relaxation rate $\gamma_\ell$ sets the linewidth of each mode's spectral feature, the fast-decaying modes ($\gamma_\ell \sim \Gamma_{\perp,\max}$) produce wide, overlapping line shapes that cannot be resolved as individual peaks, forming a smooth, broad background; in contrast, the slowly decaying modes ($\gamma_\ell < \Gamma_{\perp,\max}$) produce line shapes that are narrower than $\Gamma_{\perp,\max}$ and that are clearly resolved on top of the background at $\omega =0$. 

Although the generalized Rabi couplings are finite, resolved sideband peaks at nonzero frequencies do not exist. This can be explained as follows. Because the dissipative rates at $y \gtrsim 1$ exceed the dressed-state transition frequencies, dissipation damps the coherences faster than the coherent drive can sustain oscillations between states. Consequently, the dynamical modes are overdamped and their oscillations are suppressed, so that the oscillation frequencies $\upsilon_\ell$ of all modes, i.e., the fast- and slow-decaying modes, are close to zero. Since the oscillation frequency $\upsilon_{\ell}$ sets the position of the $\ell$th feature, this implies that the spectral feature is centered near $\omega = 0$ and that sideband peaks are not visible.

\begin{figure*}
\centering
\includegraphics{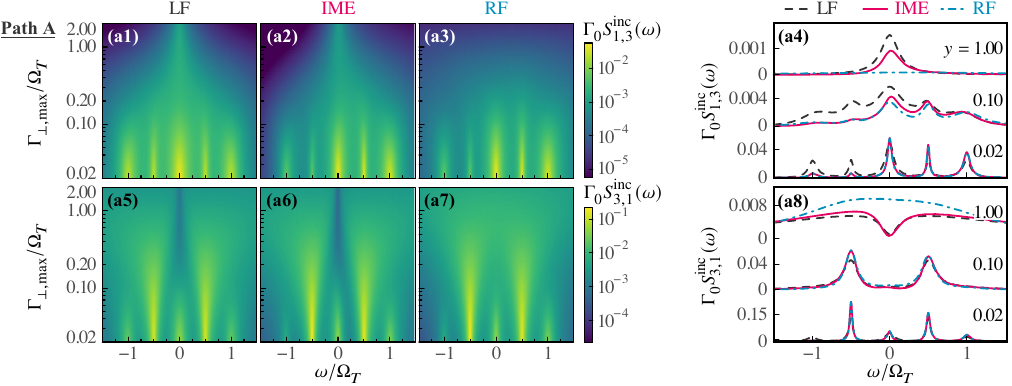}
\caption{Same layout as in Fig.~\ref{fig:transition}(a1)--(a8),
but for $\Omega_T/\kappa_{\min}\approx0.2812$ instead of $0.0889$ as in Fig.~\ref{fig:transition}.
All other parameters are the same as in Fig.~\ref{fig:transition}.}
\label{fig:transition2}
\end{figure*}

The central feature differs starkly between the emission $S^{\mathrm{inc}}_{1,3}(\omega)$ and absorption $S^{\mathrm{inc}}_{3,1}(\omega)$ spectra. Although the IME emission and absorption spectra are governed by the same generator, the transition operators $\sigma_{1,3}$ and $\sigma_{3,1}$ ``assign'' distinct spectral weights (namely, $C_\ell^{1,3}$ and $C_\ell^{3,1}$) to the dynamical modes associated with the eigenmatrix pairs $R_\ell$ and $L_\ell$ [see Eq.~\eqref{eqn:spectralweights}]. In the spectrum decomposition [see Eq.~\eqref{eqn:spectrum_modes}], the real parts $\Re[C_\ell^{\eta,\upsilon}]$ determine the amplitude and sign of each mode's Lorentzian contribution, which can be negative even though the complete spectrum is nonnegative. When calculating the emission spectrum $S^{\mathrm{inc}}_{1,3}(\omega)$, the slow-decaying modes carry a net positive weight ($\Re[C_\ell^{1,3}] > 0$) and add to the broad background to form a pronounced central maximum [see Figs.~\ref{fig:transition}(a4), (b4), and~(c4)]. When calculating the absorption spectrum $S^{\mathrm{inc}}_{3,1}(\omega)$, the slow-decaying modes carry a net negative weight ($\Re[C_\ell^{3,1}] < 0$) that gets subtracted from the positive background to produce a sharp central dip near $\omega = 0$ [see Figs.~\ref{fig:transition}(a8), (b8), and~(c8)].

The discussion can be summarized as follows. The increasing disagreement between the generalized RF and IME spectra with increasing $y$ for all three paths stems from the breakdown of the secular approximation. While the decoupled sector structure of Eqs.~\eqref{eqn:R_population}--\eqref{eqn:coherence_weights} accurately captures the distinct peaks at small $y$, which can be attributed to transitions between the dressed states, it is this decoupled sector structure that prevents the generalized RF approach to capture the nonsecular population--coherence mixing that generates the central emission maximum and absorption dip at larger $y$. The IME, which accounts for the coupling between the population and coherence sectors, captures the physics accurately also at larger $y$, provided the density matrix is semidefinite.

Compared with the two-level case, the three-level system introduces additional independently tunable drive and bath-coupling ratios. Consequently, $\Omega_T/\kappa_{\min}$ alone does not---as it would be in the two-level case~\cite{shavit2019,basak2026}---determine the accuracy of the LF approach: the influence of the ratio $\Omega_T/\kappa_{\min}$ must be considered together with the detuning, drive ratio, relative transverse and longitudinal coupling strengths, and the spectral channel that is being probed. Such an analysis can be carried out by carefully analyzing the structure of the LF generator and performing a perturbative analysis in limiting cases to connect the IME and LF generators.

To illustrate the interplay of the parameters, Fig.~\ref{fig:transition2} considers Path A spectra at $\Omega_T/\kappa_{\min}\approx0.2812$, which is larger by a factor of about $3.2$ than the value $\Omega_T/\kappa_{\min}\approx0.0889$ shown in Figs.~\ref{fig:transition}(a1)--(a8), as an example. Because $\Omega_T/\kappa_{\min}$ measures the drive-induced dressed-state frequency shift relative to the bath spectral width, increasing $\Omega_T/\kappa_{\min}$ primarily affects the accuracy of the LF approach, whereas the agreement (or disagreement) between the generalized RF and IME spectra remains similar to that in Figs.~\ref{fig:transition}(a1)--(a8). One can see that at the larger value of $\Omega_T/\kappa_{\min}$, when compared with Figs.~\ref{fig:transition}(a1)--(a8), the main contrasts persist: the LF approach misrepresents the relative weights and line shape of the emission spectrum $S^{\mathrm{inc}}_{1,3}(\omega)$, while it reproduces the central dip in the absorption spectrum $S^{\mathrm{inc}}_{3,1}(\omega)$ much more closely. We thus see notable differences between the LF and IME spectra in both Figs.~\ref{fig:transition} and~\ref{fig:transition2}. The reason the LF approach captures certain features while missing others can be understood in a complementary manner to the generalized RF approach. Unlike the generalized RF approach, the LF approach retains the nonsecular couplings that mix the population and coherence sectors at large $y$, which allows the LF approach to qualitatively capture the central emission maximum and absorption dip at $y=1$ [Figs.~\ref{fig:transition}(a4), (a8), (b4), (b8), (c4), and~(c8), and Figs.~\ref{fig:transition2}(a4) and~(a8)]. However, the LF approach evaluates the bath response at the bare atomic frequencies rather than accounting for the drive-induced dressed-state frequency shifts $E_n - E_m$. At $T_b = 0$, this bare-frequency sampling fails to suppress energetically upward longitudinal transitions between dressed states, altering the steady-state dressed populations relative to the IME; this distorts the sideband asymmetry at small $y$ and overpopulates the bright states (overestimating the emission intensity) at both small and large $y$.

\subsection{Quantifying deviations between different\texorpdfstring{\\}{ }master-equation approaches}
Section~\ref{subsec:specsign} shows that the spectra obtained within the LF and generalized RF approaches can deviate from each other and from the spectra obtained within the IME framework. This subsection quantifies the differences across a broad parameter range. Our aim is to identify parameter combinations for which the LF or generalized RF approaches reproduce the IME spectra and those for which both deviate from the IME spectra. This comparison determines when approximations can be made and when the drive-dependent dissipation rates and couplings between transitions that are accounted for by the IME must be included.

To track where the LF and generalized RF approaches capture the correct physics and where they break down, we compare the incoherent LF and generalized RF spectra $S^{\rm inc}_{\eta,\upsilon}(\omega)$ for the transition $(\eta,\upsilon)$ with the IME reference $S^{\mathrm{inc,IME}}_{\eta,\upsilon}(\omega)$ using the spectral projection metric
\begin{align}
D
&=
\frac{
\textstyle\int\limits_{-\infty}^{\infty}d\omega\,
S^{\rm inc}_{\eta,\upsilon}(\omega)
S^{\mathrm{inc,IME}}_{\eta,\upsilon}(\omega)
}{
\textstyle\int\limits_{-\infty}^{\infty}d\omega\,
\left[S^{\mathrm{inc,IME}}_{\eta,\upsilon}(\omega)\right]^2
}\,.
\label{eqn:spectral_projection}
\end{align}
This overlap yields $D=1$ if $S^{\rm inc}_{\eta,\upsilon}(\omega)$ is equal to $S^{\mathrm{inc,IME}}_{\eta,\upsilon}(\omega)$. Values above or below one flag spectra for which the LF and generalized RF approaches, on average, overestimate or underestimate the IME spectral weights, respectively. Importantly, a value of $D=1$ does not guarantee that the two spectra agree for all $\omega$, as competing errors in line shape, peak weight, and asymmetry can cancel each other. For each path defined in Table~\ref{tab:paths}, we vary---as in the previous section---$\Gamma_{\perp,\max}/\Omega_T$ at fixed $\Gamma_{\parallel,\max}/\Omega_T=0.01$, $0.1$, or $0.2$; in addition, we vary $\Omega_T/\kappa_{\min}$ (recall that Figs.~\ref{fig:transition} and \ref{fig:transition2} considered $\Omega_T/\kappa_{\min}\approx 0.0889$ and $0.2812$, respectively). 

\begin{figure}
\centering
\includegraphics{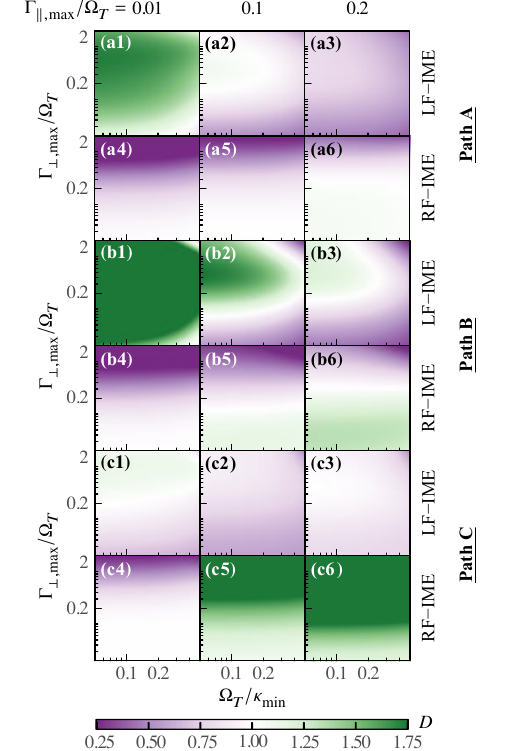}
\caption{
Spectral projection $D$ [Eq.~\eqref{eqn:spectral_projection}], 
which quantifies deviations of the LF and generalized RF approaches from the IME, 
for the incoherent emission spectrum $S^{\mathrm{inc}}_{1,3}(\omega)$ along Paths A [(a1)--(a6)], B [(b1)--(b6)], and C [(c1)--(c6)].
Panels 1--3 compare the LF with the IME spectrum, and panels 4--6 compare the generalized RF with the IME spectrum.
The first, second, and third columns correspond to $\Gamma_{\parallel,\max}/\Omega_T=0.01$, $0.1$, and $0.2$, respectively.
The magnitude and direction of the spectral projection depend on the dephasing and the parameter path.
The path constraints and bath parameters are given in Table~\ref{tab:paths}.\vspace{-1.5em}}
\label{fig:deviation}
\end{figure}

Figure~\ref{fig:deviation} shows the projection metric $D$ for the emission spectrum $S^{\rm inc}_{1,3}(\omega)$ for Paths A [(a1)--(a6)], B [(b1)--(b6)], and C [(c1)--(c6)]. For Path A, the generalized RF approach [Fig.~\ref{fig:deviation}(a4)--(a6)] performs reliably in the weak transverse decay regime (small $\Gamma_{\perp,\max}/\Omega_T$), where $D$ is close to one. However, as the transverse decay increases, $D$ deviates from one due to the breakdown of the secular approximation. The breakdown depends appreciably on the longitudinal dephasing: $D$ drops substantially below one for weak dephasing ($\Gamma_{\parallel,\max}/\Omega_T=0.01$) but the deficit reduces for stronger dephasing ($\Gamma_{\parallel,\max}/\Omega_T=0.2$). For intermediate dephasing strength ($\Gamma_{\parallel,\max}/\Omega_T=0.1$), $D$ is approximately equal to one for all $\Omega_T/\kappa_{\rm min}$ considered.

The LF approach [Fig.~\ref{fig:deviation}(a1)--(a3)], by comparison, shows that $D$ exceeds one over a substantial region for weak longitudinal coupling $\Gamma_{\parallel,\max}/\Omega_T=0.01$. However, as the longitudinal coupling $\Gamma_{\parallel,\max}/\Omega_T$ increases, $D$ is smaller than one for most of the parameters considered. As discussed in Sec.~\ref{subsec:specsign}, the LF approach's regime of validity is limited to relatively small $\Omega_T/\kappa_{\min}$, but this ratio is not the only factor governing LF--IME agreement. As $\Omega_T/\kappa_{\min}$ increases, drive-induced frequency shifts relative to the bath widths are amplified, reflecting the fact that the LF approach uses the bare transition rates. Furthermore, the LF approach does not capture the longitudinal dephasing accurately: by sampling the bath response only at zero frequency, the LF approach misses the energy differences between the dressed states and includes energetically forbidden negative-frequency contributions at zero temperature. Consequently, the parameter regimes where the LF approach is reliable and where not depends quite strongly on the ratio between the longitudinal and transverse decay.

Figures~\ref{fig:deviation}(b) and \ref{fig:deviation}(c) show analogous spectral projections for Paths B and C, respectively. For Path B, the LF approach yields a projection notably above one for nearly the entire parameter regime at $\Gamma_{\parallel,\max}/\Omega_T=0.01$ [Fig.~\ref{fig:deviation}(b1)], with the spectral projection being closer to one as the dephasing $\Gamma_{\parallel,\max}$ increases [Figs.~\ref{fig:deviation}(b2)--(b3)]. Path C exhibits a different pattern: at the two larger dephasing strengths, namely at $\Gamma_{\parallel,\max}/\Omega_T=0.1$ and $0.2$, the projection metric for the generalized RF approach is above one $(D>1)$ for a large parameter regime [Figs.~\ref{fig:deviation}(c5) and~(c6)]. The comparisons presented show that changing the microscopic parameters alters the location and the direction (i.e., toward excess or deficit) of the spectral projection metric. 

The maps of the spectral projection metric 
indicate the limitations of the LF and generalized RF approaches. They delineate the broad parameter regimes where the LF and generalized RF approaches break down, directly complementing the individual spectral cuts (examples are shown in Figs.~\ref{fig:transition} and \ref{fig:transition2}), which reveal at which frequencies spectral features are lost. 

As discussed in Sec.~\ref{sec:thesystem} the IME does not, just as the Bloch--Redfield master equation, generally guarantee complete positivity. To quantify the negativity of the density matrix, we define the negativity 
\begin{align} 
\mathcal N(\rho) = \sum_n \max\{0,-p_n\}\,, 
\label{eqn:density_negativity} 
\end{align} 
where $p_n$ are the eigenvalues of the density matrix $\rho$. The negativity $\mathcal{N}(\rho)$ vanishes for a semidefinite density matrix. For a non-semidefinite $\rho$, the smaller $\mathcal N(\rho)$ is, the more reliable the IME. We note that the negativity depends only on the properties of the density matrix. This means that the analysis presented in this section is not only relevant for the spectra considered in this work but also for other observables.

\begin{figure}
\centering
\includegraphics{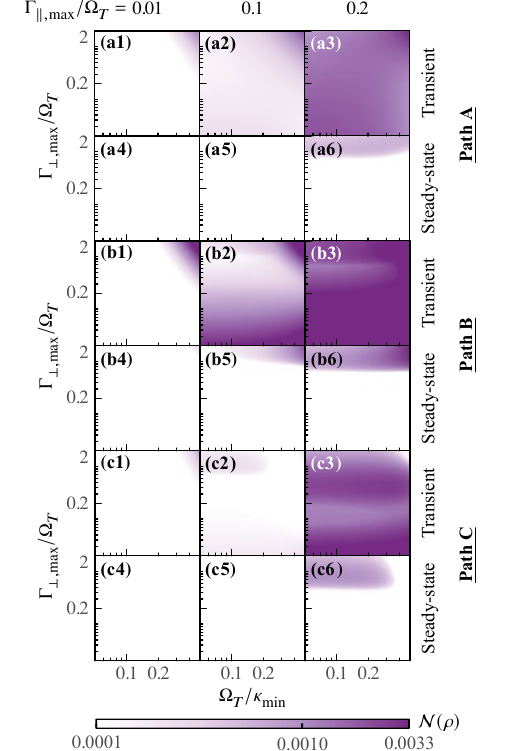}
\caption{Negativity $\mathcal N(\rho)$ [Eq.~\eqref{eqn:density_negativity}] of the 
IME density matrix for Paths A [(a1)--(a6)], B [(b1)--(b6)], and C [(c1)--(c6)]. For each path, the panels labeled 1--3 show the maximum transient negativity, $\max_{t\in T}[\mathcal N(\rho(t))]$, for the evolution from the initial state $\rho(0)=\ket{1}\bra{1}$, where $ T$ denotes the propagation time. The panels labeled 4--6 show the steady-state negativity $\mathcal N(\rho_{\mathrm{ss}})$. The first, second, and third columns correspond to $\Gamma_{\parallel,\max}/\Omega_T=0.01$, $0.1$, and $0.2$, respectively. The transient negativity is larger than the steady-state negativity. The parameter paths are the same as in Fig.~\ref{fig:deviation}.}
\label{fig:negativity}
\end{figure}

Figure~\ref{fig:negativity} shows the maximum negativity $\max_{t \in T} [{\mathcal{N}}(\rho(t))]$ over the sampled time evolution ($0<t<T$) and the steady-state negativity ${\mathcal{N}}(\rho_{ss})$ for all three paths. For the parameter combinations considered, the transient and steady-state negativities are smaller than $3.7 \times 10^{-2}$ and $3.4 \times 10^{-3}$, respectively, indicating that the IME results in quantitatively accurate results for most of the parameter space considered. 
The finite negativity at stronger dissipation may indicate that the system is approaching the limits of the Born--Markov approximation.

For Path A [Figs.~\ref{fig:negativity}(a1)--(a6)], the largest steady-state negativity $\mathcal{N}(\rho_{\mathrm{ss}})$ value of $5.75\times10^{-4}$ is seen for $\Omega_T/\kappa_{\min}=0.5$, $\Gamma_{\perp,\max}/\Omega_T=2$, and $\Gamma_{\parallel,\max}/\Omega_T=0.2$. The steady-state negativity is negligible for the vast majority of the parameter space, with minor positivity violations existing only at comparatively large dephasing and transverse decay rates. For $\Gamma_{\parallel,\max}/\Omega_T=0.1$, e.g., the transient negativity is substantially smaller than at $0.2$. Crucially, the spectral characteristics highlighted in the previous subsection, including the deviations between the three different master equation approaches, exist already at $\Gamma_{\parallel,\max}/\Omega_T=0.01$, i.e., for parameter combinations for which the negativity is extremely small or nonexistent. Figures~\ref{fig:negativity}(b) and \ref{fig:negativity}(c) show the IME density-matrix negativity for Paths B and C, respectively. As for Path A, the transient negativity is larger than the steady-state negativity, particularly at larger dephasing. For Paths B and C, the largest steady-state negativities are approximately $3.4 \times10^{-3}$ at $(\Omega_T/\kappa_{\min},\Gamma_{\perp,\text{max}}/\Omega_T) \approx(0.500,1.867)$ and $9.5\times10^{-4}$ at $(0.0975,0.796)$, respectively. Both maxima occur at $\Gamma_{\parallel,\text{max}}/\Omega_T=0.2$.

The frequency-resolved spectra (presented in Sec.~\ref{subsec:specsign}) and spectral projection metric maps (presented in this subsection) identify parameter regimes in which the LF or generalized RF approximations adequately reproduce the IME spectra and those in which the drive-dependent dissipation rates and transition couplings retained by the IME are needed. The negativity analysis complements the analysis of the spectra. If the negativity is zero or extremely small, the results obtained by the IME can be trusted. If the negativity is finite, caution needs to be exercised. Specifically, if the negativity is appreciable, closer examination of the origin and magnitude of the positivity violations is needed, including an analysis over which time scales positivity violations exist and whether they persist in the steady state. If these violations reflect a breakdown of the Born or Markov approximation, the positivity of the LF and generalized RF density matrices, which follows from the Lindblad form of the corresponding master equations, does not establish their reliability, since all three descriptions are derived by making the Born and Markov approximations. Depending on the origin of the breakdown, regimes with non-negligible positivity violations may require a treatment that retains bath memory, goes beyond the weak-coupling approximation, or incorporates both effects.

\section{\label{sec:summary}Summary and outlook}
In this work, we developed a unified master-equation framework for arbitrarily driven dissipative $N$-level quantum systems within the Born--Markov approximation. By formulating the open-system dynamics in the eigenbasis of a Lewis--Riesenfeld invariant, our approach incorporates the full driven system evolution when constructing the dissipators. Within the unified framework [Eq.~\eqref{eqn:unifiedgenME}], the invariant-based master equation (IME), the laboratory-frame (LF) master equation, and the generalized rotating-frame (RF) master equation emerge on an equal footing, distinguished by how their dissipation coefficients sample the bath spectral density and whether secular approximations are imposed on transitions between the eigenstates of the invariant. In the appropriate limits, the framework naturally recovers standard GKLS generators as well as a previously established generalized master equation for a periodically driven two-level system~\cite{basak2026}.

Specializing to a driven three-level $\Lambda$ system coupled to a Lorentzian thermal bath, we demonstrated that the unified formalism applies across qualitatively distinct driving regimes---constant-Rabi-coupling periodic driving, two-frequency Rabi-coupling-modulated protocols, and nonperiodic chirped ramps---without relying on Floquet periodicity or adiabatic separation of timescales. Comparisons of the population dynamics and the complex-plane coherence trajectories reveal that the different master equations show qualitatively different dynamics for parameter combinations for which the IME density matrix is positive semidefinite.

To establish an experimentally accessible signature of these dissipative effects, we analyzed the incoherent fluorescence spectra $S^{\rm inc}_{1,3}$ and $S^{\rm inc}_{3,1}$ for a system Hamiltonian that is stationary in the rotating frame. By decomposing the spectra into spectral weights that are linked to the dynamical modes of the Liouvillian, we traced the failure of the conventional master equations to two distinct physical mechanisms:
\begin{enumerate}
    \item \emph{Inaccurate bath-frequency sampling (breakdown of the LF approach):} Because the LF treatment evaluates bath correlation functions strictly at bare atomic transition frequencies, it neglects drive-induced dressed frequency shifts. For a zero-temperature bath and for longitudinal dephasing processes, sampling at zero bare frequency includes energetically forbidden negative-frequency transitions. Consequently, the LF approach distorts the balance between Lorentzian weights and dispersive asymmetries.
    \item \emph{Omission of nonresonant dissipative couplings (breakdown of the generalized RF approach):} Enforcing the secular approximation in the dressed basis restricts the spectral weights $C_\ell^{\eta,\upsilon}$ to be strictly real, artificially enforcing purely Lorentzian line shapes and eliminating dispersive peak asymmetries. As transverse decay increases relative to the generalized Rabi coupling, neglecting nonresonant couplings between dressed transitions causes the generalized RF description to miss the narrow central emission peak in $S^{\rm inc}_{1,3}$ and the destructive-interference dip in the absorption spectrum $S^{\rm inc}_{3,1}$.
\end{enumerate}
By calculating the projection metric $D$ alongside transient and steady-state density-matrix negativity $\mathcal{N}(\rho)$ for several parameter paths, we delineated the regimes where conventional approximations suffice and where the full IME is required. Importantly, our positivity diagnostics confirm that the differences between spectra calculated with different frameworks exist in regimes where the reduced density matrix is positive semidefinite throughout the evolution.

Looking ahead, our results provide both a theoretical foundation and a practical spectroscopic roadmap for modeling and controlling driven multilevel open quantum systems. Experimentally, the predicted spectral asymmetries and central interference features can be directly probed in platforms featuring tunable, structured baths---such as superconducting qutrits coupled to microwave photonic crystals or waveguides~\cite{Harrington2019}, surface-acoustic-wave resonators~\cite{Kitzman2023}, semiconductor quantum dots in nanocavities~\cite{ates2009}, and programmable trapped-ion simulators~\cite{Sun2025}. In such engineered reservoirs, where spectral densities vary appreciably over the scale of the Rabi splitting, accurate frequency sampling via the IME becomes essential.

From a quantum-control perspective, formulating the dissipator directly in the Lewis--Riesenfeld invariant basis opens natural avenues for inverse engineering open-system dynamics~\cite{wu2022,boubakour2025}. Since the auxiliary equations that govern the invariant link the control parameters directly to the time-dependent dissipation rates, shortcut-to-adiabaticity ~\cite{Li2021,Zhao2004,Kang2017,jin2025} and optimal-control protocols can be integrated into the IME to actively suppress decoherence or accelerate dissipative state preparation during nonadiabatic gate operations. Future theoretical extensions could generalize the explicit invariant construction to closed-loop multilevel topologies (such as $\Delta$ systems), where time-independent rotating frames do not exist, even when the drive is constant~\cite{Paing2025}, or combine the invariant propagator with non-Markovian bath kernels to explore the interplay of bath memory and driving with nonnegligible chirp.

\begin{acknowledgments}
This work is supported by the W.~M.~Keck Foundation. This material is also based upon work supported by the Air Force Office of Scientific Research under Award No.~FA9550-24-1-0106. 
The computing for this project was performed at the OU Supercomputing Center for Education \& Research (OSCER) at the University of Oklahoma (OU).
\end{acknowledgments}

\onecolumngrid
\appendix
\section{\label{sec:appendA}Bloch--Redfield master equation}
The Bloch--Redfield master equation provides the starting point for the derivations below. We define the interaction picture using the uncoupled system--bath propagator
\begin{align}
    U_T(t) = U_s(t) \otimes U_b(t)\,.\qquad\;\;
    \label{eqn:freepropagator}
\end{align}
For a general time-dependent system Hamiltonian, we have
\begin{align}
    U_s(t) &= \mathcal{T}\exp[-i\textstyle\int\limits_{0}^{t} H_s(\tau) d\tau ]
\intertext{and}
    U_b(t) &= e^{-iH_b t}\,,
\end{align}
where $\mathcal{T}$ denotes the time-ordering operator.
For a time-independent system Hamiltonian, the propagator simplifies to $U_s(t) = e^{-iH_s t}$. The propagator $U_T(t)$ transforms the total Hamiltonian from the Schr\"odinger picture to the interaction picture according to $[U_T(t)]^{\Dag} H(t) U_T(t) - i [U_T(t)]^{\Dag} \frac{d}{dt} U_T(t)$. Referring to the indices $\eta,\upsilon$ collectively as $\bar{\upsilon}$ for notational convenience and defining $B_{\bar{\upsilon}} = \sum\limits_{k}g_{\bar{\upsilon},k}\left(b_k+b^{\Dag}_k\right)$ and $S_{\bar{\upsilon}} = \sigma_{\eta,\upsilon}$, we write the interaction Hamiltonian [Eq.~\eqref{eqn:systembathintHamiltonian}] as $H_I = \sum\limits_{\bar{\upsilon}} S_{\bar{\upsilon}} \otimes B_{\bar{\upsilon}}$, where $S_{\bar{\upsilon}}$ and $B_{\bar{\upsilon}}$ act on the system and bath, respectively. The index $\bar{\upsilon}$ labels the coupling channels, including both transverse and longitudinal couplings. The total Hamiltonian in the interaction picture is 
\begin{align}
    \tilde{H} = \sum\limits_{\bar{\upsilon}} \tilde{S}_{\bar{\upsilon}}(t) \otimes \tilde{B}_{\bar{\upsilon}}(t)\,,
\end{align}
where $\tilde{S}_{\bar{\upsilon}}(t) = [U_s(t)]^{\Dag} S_{\bar{\upsilon}} U_s(t)$ and $\tilde{B}_{\bar{\upsilon}}(t) = [U_b(t)]^{\Dag} B_{\bar{\upsilon}} U_b(t)$.
Substituting this interaction Hamiltonian into the Bloch--Redfield master equation and simplifying gives 
\begin{align}
     \dfrac{d}{dt}\tilde{\rho}(t) = \sum\limits_{\bar{\upsilon},\,\bar{\upsilon}'} \int\limits_0^{\infty} d\tau \bigg\{&\left[\tilde{S}_{\bar{\upsilon}'}(t-\tau)\tilde{\rho}(t) \tilde{S}_{\bar{\upsilon}}(t)-\tilde{S}_{\bar{\upsilon}}(t)\tilde{S}_{\bar{\upsilon}'}(t-\tau)\tilde{\rho}(t) \right]\Lambda_{b,\bar{\upsilon},\bar{\upsilon}'}(\tau) \notag\\&+ \left[ \tilde{S}_{\bar{\upsilon}}(t)\tilde{\rho}(t) \tilde{S}_{\bar{\upsilon}'}(t-\tau)-\tilde{\rho}(t)\tilde{S}_{\bar{\upsilon}'}(t-\tau) \tilde{S}_{\bar{\upsilon}}(t) \right]\Lambda^{*}_{b,\bar{\upsilon},\bar{\upsilon}'}(\tau)\bigg\}\,,
\label{eq_BR}
\end{align}
where $\tilde{\rho}(t)$ is the density matrix of the driven $N$-level system in the interaction picture and $\Lambda_{b,\bar{\upsilon},\bar{\upsilon}'}(\tau) = \text{Tr}_b\{\tilde{B}_{\bar{\upsilon}}(\tau)\tilde{B}_{\bar{\upsilon}'}(0)\rho_b\}$, with the bath density matrix $\rho_b$ being unchanged by the transformation between the Schr\"odinger and interaction pictures. Equation~\eqref{eq_BR} serves as the starting point for the IME, LF, and generalized RF derivations in Appendices~\ref{sec:appendB}, \ref{sec:appendD}, and \ref{sec:appendE}, respectively.

\section{\label{sec:appendB}Invariant-based master equation}
Starting from the Bloch--Redfield master equation, Eq.~\eqref{eq_BR}, we express the system propagator using the invariant eigenstates [Eq.~\eqref{eqn:dynamicalInv}] and the Lewis--Riesenfeld phases [Eq.~\eqref{eqn:LRIphases}],
\begin{align}
    U_s(t) = \sum\limits_n e^{i\alpha_n(t)}\ket{\mu_n(t)}\bra{\mu_n(0)}\,.
\end{align}
The interaction-picture Hamiltonian is
\begin{align}
    \tilde{H}_I(t) = \sum\limits_{\eta, \upsilon} \left[\sum\limits_{m,n}\left( \xi_{m,n,\eta,\upsilon}(t)e^{+i\theta_{m,n,\eta,\upsilon}(t)}\right)F_{m,n}(0)\right]\otimes \left[\sum\limits_k g_{\eta,\upsilon,k} \left(e^{-i\omega_k t }b_k+e^{+i\omega_k t }b^{\Dag}_k\right)\right]\,.
\end{align}
Because $H_I$ is Hermitian, the interaction-picture Hamiltonian can equivalently be written as
\begin{align}
    \tilde{H}_I(t) = \sum\limits_{\eta, \upsilon} \left[\sum\limits_{m,n}\left( \xi^*_{m,n,\eta,\upsilon}(t)e^{-i\theta_{m,n,\eta,\upsilon}(t)}\right)F^{\Dag}_{m,n}(0)\right] \otimes \left[\sum\limits_k g^*_{\eta,\upsilon,k} \left(e^{-i\omega_k t }b_k+e^{+i\omega_k t }b^{\Dag}_k\right)\right]\,.
\end{align}
Substituting these interaction-picture operators into the Bloch--Redfield master equation [Eq.~\eqref{eq_BR}] and using Eqs.~\eqref{eqn:thetaexpand} and \eqref{eqn:instfreq} gives
\begin{align}
     \dfrac{d}{dt}\tilde{\rho}_s(t) = \sum\limits_{\substack{m\hphantom{'},\,n\hphantom{'}\\m',\,n'}} &\left[F_{m,n}(0)\tilde{\rho}_s(t) F^{\Dag}_{m',n'}(0) - F^{\Dag}_{m',n'}(0)F_{m,n}(0)\tilde{\rho}_s(t)\right]\nonumber 
     \\\times &\left\{\sum\limits_{\substack{\eta\hphantom{'},\,\upsilon\hphantom{'},\\ \eta',\,\upsilon'}}\!\left(\xi_{m',n',\eta',\upsilon'}(t)\right)^*e^{i\left[\theta_{m,n,\eta,\upsilon}(t)-\theta_{m',n',\eta',\upsilon'}(t)\right]} \int\limits_0^{\infty} \!\!d\tau \xi_{\vphantom{m',n'}m,n,\eta,\upsilon}(t-\tau)e^{-i\omega_{m,n,\eta,\upsilon}(t,\tau)\tau}\Lambda_{b,\eta,\upsilon,\eta',\upsilon'}(\tau)\right\} + \text{H.c.}\,,
\end{align}
where we have for a thermal bath
\begin{align}
    \Lambda_{b,\eta,\upsilon,\eta',\upsilon'}(\tau)  = \sum_{k} g^*_{\eta',\upsilon',k} g_{\eta,\upsilon,k} \left\{e^{+i\omega_k \tau}n_b(\omega_k) + e^{-i\omega_k \tau}\left[1+n_b(\omega_k)\right]\right\}\,.
    \label{eqn:eq_Lambb}
\end{align}
We replace the discrete bath modes with a continuum and use the spectral-density functions $J_{\eta,\upsilon,\eta',\upsilon'}(\omega)$ defined in Eqs.~\eqref{eqn:continuummodes1} and \eqref{eqn:continuummodes2}.
Using the spectral-density function, Eq.~\eqref{eqn:eq_Lambb} can equivalently be written as
\begin{align}
    \Lambda_{b,\eta,\upsilon,\eta',\upsilon'}(\tau)= \int\limits_0^{\infty} d\omega J_{\eta,\upsilon,\eta',\upsilon'}(\omega)\left\{e^{+i\omega \tau}n_b(\omega) + e^{-i\omega \tau}\left[1+n_b(\omega)\right]\right\}\,.
    \label{eqn:lambdab}
\end{align}
Switching to the Schr\"odinger picture gives Eq.~\eqref{eqn:unifiedgenME} with the dissipation coefficients given in Eq.~\eqref{eqn:disscoefficient-IME}.
No secular averaging or approximations for the ``delayed'' amplitudes and phases (i.e., the amplitudes and phases evaluated at $t-\tau$) were introduced in rewriting the Bloch--Redfield equation.

\section{\label{sec:appendC}Connection to the two-level generalized master equation}
Here, we demonstrate that for a periodically driven two-level system, the IME reduces to the generalized framework in Ref.~\cite{shavit2019}. The Hamiltonian for a two-level driven system is given by
\begin{align}
    H_s(t) = -\omega^0_{1} \sigma_{1,1}+ \dfrac{\Omega_{1,2}}{2}\left(e^{+i\omega^d_{1,2}t} + e^{-i\omega^d_{1,2}t}\right) \left(\sigma^{\Dag}_{1,2}+ \sigma_{1,2}\right)\,.
\end{align}
Switching to the rotating frame using $U(t)$ [Eq.~\eqref{eqn:Urot}] and applying the rotating-wave approximation, which requires $\omega^d_{1,2} \gg \sqrt{\Omega^2_{1,2}+\Delta^2}$, where $\Delta = \omega^0_{1}-\omega^d_{1,2}$, we obtain 
\begin{align}
    H^R_s = -\Delta \sigma_{1,1}+ \dfrac{\Omega_{1,2}}{2}\left(\sigma^{\Dag}_{1,2}+ \sigma_{1,2}\right)\,.
\end{align}
As the Hamiltonian is time independent, the dynamical invariant can be made time independent. Using the initial condition 
\begin{align}
\comm{H^R_s}{I} =0\,,
\end{align}
the Hamiltonian and the dynamical invariant in the rotating frame have the same eigenstates (the dressed states):
\begin{align}
    \ket{\mu_+}
    =&
    \sin\left(\dfrac{\beta}{2}\right)\ket{1}
    +
    \cos\left(\dfrac{\beta}{2}\right)\ket{2}\,,
    \nonumber\\
    \ket{\mu_-}
    =&
    \cos\left(\dfrac{\beta}{2}\right)\ket{1}
    -
    \sin\left(\dfrac{\beta}{2}\right)\ket{2}\,,
    \label{eqn:2levelinvarianteigenstates}
\end{align}
where $\beta = \tan^{-1}
    \left(\Omega_{1,2}/\Delta\right)$. In the rotating frame, the interaction Hamiltonian is given by 
\begin{align}
    H^R_I = \sum\limits_k\left[ g_{1,2,\,k}e^{-i\omega^d_{1,2}t}\sigma_{1,2} +  g^*_{1,2,\,k}e^{+i\omega^d_{1,2}t}\sigma^{\Dag}_{1,2} + g_{1,1,k}\left(\sigma_{2,2}-\sigma_{1,1}\right)\right]\left(b_k+b^{\Dag}_k\right)\,.
\end{align}
We find that the amplitudes of the matrix elements in the basis $\{\ket{\mu_+},\ket{\mu_-}\}$ are given by
\begin{subequations}
   \begin{align}
   \xi_{m=\pm,n=\pm, 1,2}= \begin{pmatrix}
       \dfrac{1}{2}\sin\beta & -\sin^2\left(\dfrac{\beta}{2}\right)\\[10pt]
       \cos^2\left(\dfrac{\beta}{2}\right) & -\dfrac{1}{2}\sin\beta
   \end{pmatrix}
\end{align} 
and 
   \begin{align}
   \xi_{m=\pm,n=\pm, z}= \begin{pmatrix}
       \cos\beta & -\sin\beta\\[5pt]
       -\sin\beta & -\cos\beta
   \end{pmatrix}\,,
\end{align} 
\end{subequations}
where $\xi_{m,n, z} = \bra{\mu_m}\sigma_{2,2}-\sigma_{1,1}\ket{\mu_n}$. The phases are
\begin{subequations}
\begin{align}
    \theta_{m=\pm,n=\pm, 1,2}(t) = \begin{pmatrix}
       -\omega^d_{1,2}t & \Omega_T t-\omega^d_{1,2}t  \\
       -\Omega_T t -\omega^d_{1,2}t & -\omega^d_{1,2}t
   \end{pmatrix}
\end{align}
and
\begin{align}
    \theta_{m=\pm,n=\pm, z}(t) = \begin{pmatrix}
       0 & \Omega_T t \\
       -\Omega_T t  & 0
   \end{pmatrix}\,, 
\end{align}
\end{subequations}
where $\Omega_T = \sqrt{\Delta^2 + \Omega^2_{1,2}}$ and $\theta_{m,n,z} = \alpha_n(t)-\alpha_m(t)$.

To connect our dissipation coefficients to the rate parameters of Ref.~\cite{shavit2019}, we neglect the Lamb shifts (i.e., the imaginary parts of the bath integrals) and write
\begin{subequations}
\label{eqn:twolevel-bath-rates}
\begin{align}
    \Gamma_\downarrow&=2\Re\left[\int_0^\infty d\tau\,
    e^{i\omega^d_{1,2}\tau}\Lambda_{b,1,2,1,2}(\tau)\right]
    \,,
    \label{eqn:C9a}
    \\
    \Gamma_\uparrow&=2\Re\left[\int_0^\infty d\tau\,
    e^{-i\omega^d_{1,2}\tau}\Lambda_{b,1,2,1,2}(\tau)\right]
    \,,
    \label{eqn:C9b}
    \\
    \Gamma_0^z&=2\Re\left[\int_0^\infty d\tau\,
    \Lambda_{b,1,1,1,1}(\tau)\right]
    \,,
    \label{eqn:C9c}
    \\
    \Gamma_\pm^z&=2\Re\left[\int_0^\infty d\tau\,
    e^{\pm i\Omega_T\tau}\Lambda_{b,1,1,1,1}(\tau)\right]
    \,,
    \label{eqn:C9d}
    \\
    \Gamma_\downarrow(1\pm\epsilon)&\approx2\Re\left[\int_0^\infty d\tau\,
    e^{i(\omega^d_{1,2}\pm\Omega_T)\tau}
    \Lambda_{b,1,2,1,2}(\tau)\right]
    \,,
    \label{eqn:C9e}
    \intertext{and}
    \Gamma_\uparrow(1\mp\epsilon_e)&\approx2\Re\left[\int_0^\infty d\tau\,
    e^{i(-\omega^d_{1,2}\pm\Omega_T)\tau}
    \Lambda_{b,1,2,1,2}(\tau)\right]
    \,.
    \label{eqn:C9f}
\end{align}
\end{subequations}
Here, Eqs.~\eqref{eqn:C9a}, \eqref{eqn:C9b}, \eqref{eqn:C9c}, and \eqref{eqn:C9d} define $\Gamma_{\downarrow}$, $\Gamma_{\uparrow}$, $\Gamma_{0}^{z}$, and $\Gamma_{\pm}^{z}$, respectively, whereas Eqs.~\eqref{eqn:C9e} and \eqref{eqn:C9f} approximate the transverse dissipation rates at the sideband frequencies $\pm\omega_{1,2}^{d}\pm\Omega_{T}$ via a first-order Taylor expansion in $\Omega_{T}$ about $\pm\omega_{1,2}^{d}$, as done in Ref.~\cite{shavit2019}. 

Substituting $\Lambda_{b,1,2,1,2}(\tau)$ from Eq.~\eqref{eqn:lambdab} and using $\Re\!\left[\int_{0}^{\infty} d\tau\, e^{i(\nu\mp\omega)\tau}\right] = \pi\delta(\nu\mp\omega)$ together with $\omega_{1,2}^{d} \gg \Omega_{T} > 0$, Eqs.~\eqref{eqn:C9a},~\eqref{eqn:C9b},~\eqref{eqn:C9e}, and \eqref{eqn:C9f} evaluate to
\begin{subequations}
\label{eqn:C_integrals_exact}
\begin{align}
\Gamma_{\downarrow} &= 2\pi J_{1,2}(\omega_{1,2}^{d})\left[1 + n_{b}(\omega_{1,2}^{d})\right], \label{eqn:C_Gamma_down_exact}\\
\Gamma_{\uparrow} &= 2\pi J_{1,2}(\omega_{1,2}^{d})n_{b}(\omega_{1,2}^{d}), \label{eqn:C_Gamma_up_exact}\\
2\,\Re\left[\int_{0}^{\infty} d\tau \, e^{i(\omega_{1,2}^{d}\pm\Omega_{T})\tau}\Lambda_{b,1,2,1,2}(\tau)\right] 
&= 2\pi J_{1,2}(\omega_{1,2}^{d}\pm\Omega_{T})\left[1 + n_{b}(\omega_{1,2}^{d}\pm\Omega_{T})\right], \label{eqn:C_sideband_pos}
\intertext{and}
2\,\Re\left[\int_{0}^{\infty} d\tau \, e^{i(-\omega_{1,2}^{d}\pm\Omega_{T})\tau}\Lambda_{b,1,2,1,2}(\tau)\right] 
&= 2\pi J_{1,2}(\omega_{1,2}^{d}\mp\Omega_{T})n_{b}(\omega_{1,2}^{d}\mp\Omega_{T}), \label{eqn:C_sideband_neg}
\end{align}
\end{subequations}
where $J_{1,2}(\omega)$ is the bath spectral density function. Assuming that $J_{1,2}(\omega)$ and $n_{b}(\omega)$ vary smoothly around $\omega_{1,2}^{d}$ on the scale of $\Omega_{T}$, we Taylor expand the right-hand sides of Eqs.~\eqref{eqn:C_sideband_pos} and \eqref{eqn:C_sideband_neg} to first order in $\Omega_{T}$ about $\omega_{1,2}^{d}$:
\begin{subequations}
\label{eqn:C_taylor}
\begin{align}
2\pi J_{1,2}(\omega_{1,2}^{d}\pm\Omega_{T})\left[1 + n_{b}(\omega_{1,2}^{d}\pm\Omega_{T})\right] 
&\approx \Gamma_{\downarrow} \pm 2\pi\Omega_{T} \left.\frac{d}{d\omega}\Big\{J_{1,2}(\omega)\left[1 + n_{b}(\omega)\right]\Big\}\right|_{\omega=\omega_{1,2}^{d}} 
= \Gamma_{\downarrow}(1 \pm \epsilon) \label{eqn:C_taylor_pos}
\intertext{and}
2\pi J_{1,2}(\omega_{1,2}^{d}\mp\Omega_{T})n_{b}(\omega_{1,2}^{d}\mp\Omega_{T}) 
&\approx \Gamma_{\uparrow} \mp 2\pi\Omega_{T} \left.\frac{d}{d\omega}\Big[J_{1,2}(\omega)n_{b}(\omega)\Big]\right|_{\omega=\omega_{1,2}^{d}} 
= \Gamma_{\uparrow}(1 \mp \epsilon_{e})\,. \label{eqn:C_taylor_neg}
\end{align}
\end{subequations}
Comparing Eqs.~\eqref{eqn:C_taylor_pos} and \eqref{eqn:C_taylor_neg} with Eqs.~\eqref{eqn:C9e} and \eqref{eqn:C9f}, we find 
\begin{subequations}
\label{eqn:C_eps_def}
\begin{align}
\epsilon &= \Omega_{T}\left.\frac{d}{d\omega}\ln\!\Big\{J_{1,2}(\omega)\left[1 + n_{b}(\omega)\right]\Big\}\right|_{\omega=\omega_{1,2}^{d}}\label{eqn:C_eps}
\intertext{and}
\epsilon_{e} &= \Omega_{T}\left.\frac{d}{d\omega}\ln\!\Big[J_{1,2}(\omega)n_{b}(\omega)\Big]\right|_{\omega=\omega_{1,2}^{d}}
\,. \label{eqn:C_epse}
\end{align}
\end{subequations}
Equations~\eqref{eqn:C_eps} and \eqref{eqn:C_epse} show explicitly that $\epsilon$ and $\epsilon_{e}$ are dimensionless and linear in $\Omega_{T}$. We note that the sign convention for $\epsilon_{e}$ in Eqs.~\eqref{eqn:C9f} and \eqref{eqn:C_taylor_neg} follows Eq.~(A7) of Ref.~\cite{shavit2019}.

We discard the terms that oscillate near $\omega^d_{1,2}$ or higher frequencies and retain those that oscillate at $\Omega_T$ and $2\Omega_T$. Substitution of the amplitudes and phases into the dissipation coefficients gives
\begin{subequations}
\label{eqn:twolevel-dissipation-coefficients}
\begin{align}
    \Gamma_{+,+,+,+}
    ={}&
    \frac{\sin^2\beta}{8}
    \left(\Gamma_\downarrow+\Gamma_\uparrow\right)
    +
    \frac{\cos^2\beta}{2}\Gamma_0^z\,,
    \label{eqn:twolevel-Gamma-diagonal}
    \\
    \Gamma_{\pm,\mp,\pm,\mp}
    ={}&
    \frac{(1\mp\cos\beta)^2}{8}
    \Gamma_\downarrow(1\mp\epsilon)
    +
    \frac{(1\pm\cos\beta)^2}{8}
    \Gamma_\uparrow(1\pm\epsilon_e)
    +
    \frac{\sin^2\beta}{2}\Gamma_\mp^z\,,
    \label{eqn:twolevel-Gamma-transition}
    \\
    \Gamma_{\pm,\mp,+,+}
    ={}&
    \frac{\sin\beta}{8}
    \Big[
    (\cos\beta\mp1)
    \Gamma_\downarrow(1\mp\epsilon)
    +
    (\cos\beta\pm1)
    \Gamma_\uparrow(1\pm\epsilon_e)
    -
    4\cos\beta\,\Gamma_\mp^z
    \Big]\,,
    \label{eqn:twolevel-Gamma-transition-diagonal}
    \\
    \Gamma_{+,+,\pm,\mp}
    ={}&
    \frac{\sin\beta}{8}
    \Big[
    (\cos\beta\mp1)\Gamma_\downarrow
    +
    (\cos\beta\pm1)\Gamma_\uparrow
    -
    4\cos\beta\,\Gamma_0^z
    \Big]\,,
    \label{eqn:twolevel-Gamma-diagonal-transition}
    \intertext{and}
    \Gamma_{\pm,\mp,\mp,\pm}
    ={}&
    -\frac{\sin^2\beta}{8}
    \Big[
    \Gamma_\downarrow(1\mp\epsilon)
    +
    \Gamma_\uparrow(1\pm\epsilon_e)
    -
    4\Gamma_\mp^z
    \Big]\,.
    \label{eqn:twolevel-Gamma-opposite-transitions}
\end{align}
\end{subequations}
With rows and columns ordered as $\{(+,+),(+,-),(-,+)\}$,
the dissipation coefficients can equivalently be written as
\begin{equation}
\label{eqn:twolevel-dissipation-coefficients-compact}
\boldsymbol{\Gamma}
=
\frac{1}{8}
\begin{pmatrix}
\sin\beta\,\Gamma_\downarrow
&
\sin\beta\,\Gamma_\uparrow
&
2\cos\beta\,\Gamma_0^z
\\
(\cos\beta-1)\Gamma_\downarrow(1-\epsilon)
&
(\cos\beta+1)\Gamma_\uparrow(1+\epsilon_e)
&
-2\sin\beta\,\Gamma_-^z
\\
(\cos\beta+1)\Gamma_\downarrow(1+\epsilon)
&
(\cos\beta-1)\Gamma_\uparrow(1-\epsilon_e)
&
-2\sin\beta\,\Gamma_+^z
\end{pmatrix}
\begin{pmatrix}
\sin\beta & \cos\beta-1 & \cos\beta+1
\\
\sin\beta & \cos\beta+1 & \cos\beta-1
\\
2\cos\beta & -2\sin\beta & -2\sin\beta
\end{pmatrix}\,.
\end{equation}

The remaining coefficients can be obtained by realizing that swapping the invariant eigenstates $\ket{\mu_+}$ and $\ket{\mu_-}$ reverses the signs of the corresponding complex transition amplitudes:
\begin{align}
    \Gamma_{m,n,-,-}
    =
    -\Gamma_{m,n,+,+},~
    \Gamma_{-,-,m,n}
    =
    -\Gamma_{+,+,m,n}.
    \label{eqn:twolevel-Gamma-diagonal-relations}
\end{align}
We note that $\Gamma_{\pm,\mp,+,+}$ samples the dressed frequencies (generalized Rabi coupling), whereas $\Gamma_{+,+,\pm,\mp}$ samples the drive and zero frequencies (this can be seen by realizing that the first index pair specifies the delayed operator in the bath integral).

To separate the contributions according to their oscillation frequencies, we write
\begin{align}
    \widetilde{\rho}_{m,n}(t)
    = \matrixel{\,\mu_m(0)\,}{\;\widetilde{\rho}_s(t)\,}{\,\mu_n(0)\,}
\end{align}
and suppress the argument $0$ on $F_{m,n}(0)$ for notational simplicity.
Using
\begin{align}
    F_{m,n}F_{m',n'}
    =
    \delta_{n,m'}F_{m,n'}
    \intertext{and}
    \widetilde{\rho}_{+,+}+\widetilde{\rho}_{-,-}
    =
    1,
\end{align}
the interaction-picture IME becomes
\begin{align}
    \frac{d}{dt}\widetilde{\rho}_s(t)
    =
    \mathcal{D}_0
    +
    \mathcal{D}_{\Omega_T}
    +
    \mathcal{D}_{2\Omega_T}
    +
    \mathrm{H.c.},
    \label{eqn:twolevel-IME-frequency-decomposition}
\end{align}
where $\mathcal{D}_0$, $\mathcal{D}_{\Omega_T}$, and $\mathcal{D}_{2\Omega_T}$ collect nonoscillating contributions, contributions oscillating at $\pm\Omega_T$, and contributions oscillating at $\pm2\Omega_T$, respectively,
\begin{subequations}
\label{eqn:twolevel-Ds}
\begin{align}
    &\mathcal{D}_0
    =
    -\Gamma_{-,+,-,+}
    \left[
    \widetilde{\rho}_{+,+}(F_{+,+}-F_{-,-})
    +
    \widetilde{\rho}_{+,-}F_{+,-}
    \right]
    +
    \Gamma_{+,-,+,-}
    \left[
    \widetilde{\rho}_{-,-}(F_{+,+}-F_{-,-})
    -
    \widetilde{\rho}_{-,+}F_{-,+}
    \right]-
    2\Gamma_{+,+,+,+}
    \left(
    \widetilde{\rho}_{-,+}F_{-,+}
    +
    \widetilde{\rho}_{+,-}F_{+,-}
    \right)\,,
    \label{eqn:twolevel-D0}\\
    &\mathcal{D}_{\Omega_T}
    =
    2e^{-i\Omega_Tt}
    \Gamma_{-,+,+,+}
    \widetilde{\rho}_{+,+}F_{-,+}-
    2e^{+i\Omega_Tt}
    \Gamma_{+,-,+,+}
    \widetilde{\rho}_{-,-}F_{+,-}+
    e^{+i\Omega_Tt}\Gamma_{+,+,-,+}
    \left[
    F_{+,-}
    +
    \widetilde{\rho}_{-,+}(F_{+,+}-F_{-,-})
    \right]\nonumber
    \\&\phantom{\mathcal{D}_{\Omega_T}=2e^{+i\Omega_Tt}
    \Gamma_{-,+,+,+}
    \widetilde{\rho}_{+,+}F_{-,+}-
    2e^{-i\Omega_Tt}
    \Gamma_{+,-,+,+}
    \widetilde{\rho}_{-,-}F_{+,-}}-
    e^{-i\Omega_Tt}\Gamma_{+,+,+,-}
    \left[
    F_{-,+}
    -
    \widetilde{\rho}_{+,-}(F_{+,+}-F_{-,-})
    \right]\,,
    \label{eqn:twolevel-DOmega}
    \intertext{and}
    &\mathcal{D}_{2\Omega_T}
    =
    e^{-2i\Omega_Tt}
    \Gamma_{-,+,+,-}
    \widetilde{\rho}_{+,-}F_{-,+}
    +
    e^{+2i\Omega_Tt}
    \Gamma_{+,-,-,+}
    \widetilde{\rho}_{-,+}F_{+,-}\,.
    \label{eqn:twolevel-D2Omega}
\end{align}
\end{subequations}

We are now ready to connect the resulting interaction-picture equation with Eq.~(A6) of Ref.~\cite{shavit2019}. To compare the coefficients, we continue to use $\beta$  and denote the angle $\beta$ that appears in Eq.~(A7) of Ref.~\cite{shavit2019} by $\beta_{\mathrm{Sh}}$. We identify the ``translations'' (the quantities on the left and right are used in our formulation and Ref.~\cite{shavit2019}, respectively)
\begin{align}
    \sin\beta_{\mathrm{Sh}}
    =
    \cos\beta,~   
    \cos\beta_{\mathrm{Sh}}
    =
    -\sin\beta,~   
    \omega
    =
    \Omega_T.
    \label{eqn:twolevel-angle-frequency-map}
\end{align}
The operator and density-matrix element identifications are 
\begin{subequations}
\label{eqn:twolevel-operator-density-map}
\begin{align}
    F_{+,+}-F_{-,-}
    &\longleftrightarrow
    -\sigma_z\,,
    &
    F_{-,+}
    &\longleftrightarrow
    -\sigma_+\,,
    &
    F_{+,-}
    &\longleftrightarrow
    -\sigma_-\,,
    \\
    \widetilde{\rho}_{+,+}
    &\longleftrightarrow
    u\,,
    &
    \widetilde{\rho}_{-,-}
    &\longleftrightarrow
    d\,,
    \\
    \widetilde{\rho}_{-,+}
    &\longleftrightarrow
    -x^*\,,
    &
    \widetilde{\rho}_{+,-}
    &\longleftrightarrow
    -x\,.
\end{align}
\end{subequations}
These identifications explicitly connect our invariant eigenbasis components to the rotating-frame density-matrix elements defined in Ref.~\cite{shavit2019}.

Using Eq.~\eqref{eqn:twolevel-angle-frequency-map}, the nine dissipation coefficients in Eqs.~\eqref{eqn:twolevel-dissipation-coefficients} that appear in Eqs.~\eqref{eqn:twolevel-D0}--\eqref{eqn:twolevel-D2Omega} can be written as
\begin{subequations}
\label{eqn:twolevel-mapped-gammas}
\begin{align}
    \Gamma_{-,+,-,+}
    ={}&
    \frac{(\sin\beta_{\mathrm{Sh}}+1)^2}{8}
    \Gamma_\downarrow(1+\epsilon)
    +
    \frac{(\sin\beta_{\mathrm{Sh}}-1)^2}{8}
    \Gamma_\uparrow(1-\epsilon_e)
    +
    \frac{\cos^2\beta_{\mathrm{Sh}}}{2}\Gamma_+^z,
    \\
    \Gamma_{+,-,+,-}
    ={}&
    \frac{(\sin\beta_{\mathrm{Sh}}-1)^2}{8}
    \Gamma_\downarrow(1-\epsilon)
    +
    \frac{(\sin\beta_{\mathrm{Sh}}+1)^2}{8}
    \Gamma_\uparrow(1+\epsilon_e)
    +
    \frac{\cos^2\beta_{\mathrm{Sh}}}{2}\Gamma_-^z,
    \\
    \Gamma_{+,+,+,+}
    ={}&
    \frac{1}{2}
    \left[
    \sin^2\beta_{\mathrm{Sh}}\,\Gamma_0^z
    +
    \frac{\cos^2\beta_{\mathrm{Sh}}}{4}
    \left(\Gamma_\downarrow+\Gamma_\uparrow\right)
    \right],
    \\
    \Gamma_{-,+,+,+}
    ={}&
    -\frac{\cos\beta_{\mathrm{Sh}}}{2}
    \left[
    \frac{\sin\beta_{\mathrm{Sh}}+1}{4}
    \Gamma_\downarrow(1+\epsilon)
    +
    \frac{\sin\beta_{\mathrm{Sh}}-1}{4}
    \Gamma_\uparrow(1-\epsilon_e)
    -
    \sin\beta_{\mathrm{Sh}}\,\Gamma_+^z
    \right],
    \\
    \Gamma_{+,-,+,+}
    ={}&
    -\frac{\cos\beta_{\mathrm{Sh}}}{2}
    \left[
    \frac{\sin\beta_{\mathrm{Sh}}-1}{4}
    \Gamma_\downarrow(1-\epsilon)
    +
    \frac{\sin\beta_{\mathrm{Sh}}+1}{4}
    \Gamma_\uparrow(1+\epsilon_e)
    -
    \sin\beta_{\mathrm{Sh}}\,\Gamma_-^z
    \right],
    \\
    \Gamma_{+,+,-,+}
    ={}&
    -\frac{\cos\beta_{\mathrm{Sh}}}{2}
    \left[
    \frac{\sin\beta_{\mathrm{Sh}}+1}{4}\Gamma_\downarrow
    +
    \frac{\sin\beta_{\mathrm{Sh}}-1}{4}\Gamma_\uparrow
    -
    \sin\beta_{\mathrm{Sh}}\,\Gamma_0^z
    \right],
    \\
    \Gamma_{+,+,+,-}
    ={}&
    -\frac{\cos\beta_{\mathrm{Sh}}}{2}
    \left[
    \frac{\sin\beta_{\mathrm{Sh}}-1}{4}\Gamma_\downarrow
    +
    \frac{\sin\beta_{\mathrm{Sh}}+1}{4}\Gamma_\uparrow
    -
    \sin\beta_{\mathrm{Sh}}\,\Gamma_0^z
    \right],
    \\
    \Gamma_{-,+,+,-}
    ={}&
    \frac{\sin^2\beta_{\mathrm{Sh}}-1}{8}
    \Gamma_\downarrow(1+\epsilon)
    +
    \frac{\sin^2\beta_{\mathrm{Sh}}-1}{8}
    \Gamma_\uparrow(1-\epsilon_e)
    +
    \frac{\cos^2\beta_{\mathrm{Sh}}}{2}\Gamma_+^z,
    \\
    \Gamma_{+,-,-,+}
    ={}&
    \frac{\sin^2\beta_{\mathrm{Sh}}-1}{8}
    \Gamma_\downarrow(1-\epsilon)
    +
    \frac{\sin^2\beta_{\mathrm{Sh}}-1}{8}
    \Gamma_\uparrow(1+\epsilon_e)
    +
    \frac{\cos^2\beta_{\mathrm{Sh}}}{2}\Gamma_-^z\,.
\end{align}
\end{subequations}
Substituting Eqs.~\eqref{eqn:twolevel-angle-frequency-map}, \eqref{eqn:twolevel-operator-density-map}, and \eqref{eqn:twolevel-mapped-gammas} into Eqs.~\eqref{eqn:twolevel-D0}--\eqref{eqn:twolevel-D2Omega} yields Eq.~(A7) of Ref.~\cite{shavit2019}. Together with Eq.~\eqref{eqn:twolevel-IME-frequency-decomposition}, which matches Eq.~(A6) of Ref.~\cite{shavit2019}, this establishes that the IME exactly reproduces the generalized master equation of Ref.~\cite{shavit2019} for a periodically driven two-level system. This exact reproduction is important because it benchmarks our unified framework against an established nonsecular treatment for $N=2$ while demonstrating that the invariant-based formulation naturally extends this physics—incorporating both drive-modified bath sampling and nonsecular dissipative couplings—to arbitrary multilevel systems and time-dependent, nonperiodic driving protocols.

\section{\label{sec:appendD}Laboratory-frame master equation}
In the LF approach, the driving term is neglected when constructing the dissipators but retained in the coherent evolution. In the absence of the drive, the system Hamiltonian reads
\begin{align}
    H^0_s = -\sum\limits_{\eta<N}\omega^0_{\eta} \sigma_{\eta,\eta}\,. 
\end{align}
The corresponding interaction-picture Hamiltonian reads
    \begin{align}
    \tilde{H}(t) = \sum\limits_{\eta,\,\upsilon} e^{-i\omega^0_{\eta,\,\upsilon} t}\sigma_{\eta,\upsilon} \otimes \left[\sum\limits_k  g_{\eta,\,\upsilon,\,k} \left(e^{-i\omega_k t}b_k+e^{+i\omega_k t}b^{\Dag}_k\right)\right]\,,
\end{align}
where $\omega^0_{\eta,\,\upsilon} = \omega^0_{\eta}-\omega^0_{\upsilon}$. Substituting these operators into the Bloch--Redfield equation [Eq.~\eqref{eq_BR}] gives the LF dissipators. Returning to the Schr\"odinger picture and using the full driven Hamiltonian in the coherent term, without making any secular approximations, gives
\begin{align}
     \dfrac{d}{dt}\rho(t) = -i\comm{H_s(t)}{\rho(t)} &+ \sum\limits_{\substack{\eta\hphantom{'},\,\upsilon\hphantom{'}\\ \eta',\,\upsilon'}}  \left\{\left[\sigma_{\eta,\upsilon}\rho(t) \sigma_{\upsilon',\eta'} -\sigma_{\upsilon',\eta'}\sigma_{\eta,\upsilon}\rho(t) \right]\int\limits_0^{\infty} d\tau e^{i\omega^0_{\eta,\upsilon} \tau} \Lambda_{b,\eta,\upsilon,\eta',\upsilon'}(\tau)\right\}\nonumber
     \\&+\sum\limits_{\substack{\eta\hphantom{'},\,\upsilon\hphantom{'}\\ \eta',\,\upsilon'}}  \left\{\left[\sigma_{\upsilon',\eta'}\rho(t) \sigma_{\eta,\upsilon}-\rho(t)\sigma_{\eta,\upsilon} \sigma_{\upsilon',\eta'} \right]\int\limits_0^{\infty} d\tau e^{i\omega^0_{\eta,\upsilon} \tau}\Lambda^{*}_{b,\eta,\upsilon,\eta',\upsilon'}(\tau)\right\}\,.
     \label{eqn:LFbarebasis}
\end{align}
Equation~\eqref{eqn:LFbarebasis} can be written in Lindblad form by making both a secular approximation and a uniform structureless bath approximation~\cite{mccauley2020}. In the secular approximation, terms oscillating at transition-frequency differences much larger than the decay rate are neglected. For the remaining transition frequencies, the uniform-structureless-bath approximation assumes that the bath spectral density is effectively constant. For a structureless bath characterized by~\cite{mccauley2020}
\begin{align}
    \pi J_{\eta,\upsilon}(\omega^0_{\eta,\upsilon}) &= \Gamma_{\eta,\upsilon}
\end{align}
and
\begin{align}
    \pi J_{\eta',\upsilon'}(\omega^0_{\eta',\upsilon'}) &= \Gamma_{\eta',\upsilon'}\,,
\end{align}
the cross spectral-density function becomes 
\begin{align}
    \pi J_{\eta,\upsilon,\eta',\upsilon'}(\omega^0_{\eta,\upsilon}) \approx \pi J_{\eta,\upsilon,\eta',\upsilon'}(\omega^0_{\eta',\upsilon'}) \approx \sqrt{\Gamma_{\eta,\upsilon}\Gamma_{\eta',\upsilon'}}\,.
\end{align}
Since evaluating Lamb shifts requires that we specialize to a particular form of the spectral function, we neglect Lamb shifts in what follows. Using 
\begin{align}
    \pi J_{\eta,\upsilon,\eta',\upsilon'}(\omega^0_{\eta,\upsilon}) = \Re\left[\int\limits_0^{\infty} d\tau e^{i\omega^0_{\eta,\upsilon} \tau} \Lambda_{b,\eta,\upsilon,\eta',\upsilon'}(\tau)\right]
\end{align}
and defining the composite jump operators
\begin{align}
    \sigma_{\eta,\upsilon,\eta',\upsilon'} = \sqrt{\Gamma_{\eta,\upsilon}} \sigma_{\eta,\upsilon} + \sqrt{\Gamma_{\eta',\upsilon'}} \sigma_{\eta',\upsilon'}\,,
\end{align}
the master equation reduces to Lindblad form~\cite{mccauley2020}
\begin{align}
    \dfrac{d}{dt}\rho(t) = -i\comm{H_s(t)}{\rho(t)} &+ \sum\limits_{\substack{\eta\hphantom{'},\,\upsilon\hphantom{'}\\ \eta',\,\upsilon'}} \left[2\sigma_{\eta,\upsilon,\eta',\upsilon'}\rho(t) \sigma^{\Dag}_{\eta,\upsilon,\eta',\upsilon'} -\acomm{\sigma^{\Dag}_{\eta,\upsilon,\eta',\upsilon'}\sigma_{\eta,\upsilon,\eta',\upsilon'}}{\rho(t)}\right]\,.
\end{align}
In what follows, we proceed with the LF master equation given in Eq.~\eqref{eqn:LFbarebasis} and express its transition operators $\sigma_{\eta,\upsilon}$ in the invariant eigenbasis to transform it to the unified master-equation form.
Rewriting the transition operators in the eigenbasis of the invariant
\begin{align}
    \sigma_{\eta,\upsilon} = \sum_{m,n}F_{m,n}(t)\xi_{m,n, \eta,\upsilon}(t)e^{i\varsigma_{m,n,\eta,\upsilon}(t)}
\end{align}
gives Eq.~\eqref{eqn:unifiedgenME} with the dissipation coefficients given in Eq.~\eqref{eqn:disscoefficient-LF}.

\section{\label{sec:appendE}Generalized rotating-frame master equation}

In its standard formulation, the rotating-frame master equation is derived in a frame in which the system Hamiltonian is time independent~\cite{shavit2019}. 
Constructing such a frame generally requires application of the rotating-wave approximation, which assumes that the drive frequencies are much larger than the characteristic rotating-frame frequency scales, which are set by the generalized Rabi couplings. 
For an $N$-level system with $N \leq 3$ and time-independent Rabi couplings, the rotating-wave approximation guarantees that such a Hamiltonian exists. 
For $N > 3$, if the number of Rabi couplings exceeds $N$, additional constraints on the transition and drive frequencies are generally required for the rotating-frame Hamiltonian to be time independent~\cite{Paing2025}. 
For arbitrary time-dependent driving, a time-independent rotating-frame Hamiltonian is typically unavailable, and the standard rotating-frame master equation cannot be applied.

In general, the most convenient rotating frame is obtained by transforming the full Hamiltonian under a time-dependent unitary rotation $U(t) \otimes \mathbb{I}_b$ [e.g., Eq.~\eqref{eqn:Urot}] chosen to render the system Hamiltonian time independent; here, $\mathbb{I}_b$ denotes the identity operator on the bath Hilbert space. While the rotation does not affect the bath Hamiltonian, it does affect the system--bath interaction Hamiltonian. The operator $U(t)$ [e.g., Eq.~\eqref{eqn:Urot}] dresses the system operators in the system--bath interaction Hamiltonian [e.g., Eq.~\eqref{eqn:systembathintHamiltonianrot}] with the corresponding rotating-frame phase factors. The conventional RF (Lindblad) master equation is obtained by 
(a) transforming to the eigenbasis of the time-independent Hamiltonian in the rotating frame, 
(b) using Eq.~\eqref{eqn:freepropagator} to switch from the Schr\"odinger picture to the interaction picture,
(c) applying the Born--Markov approximation, and
(d) applying a secular approximation that requires that the generalized Rabi couplings are much larger than the decay rates. 

To place the RF approach within the unified master-equation framework [Eq.~\eqref{eqn:unifiedgenME}] and extend it to arbitrarily driven $N$-level systems, we formulate the derivation in the eigenbasis of the dynamical invariant $I(t)$. 
Doing so serves two purposes. 
First, when a rotating frame exists in which the system Hamiltonian is time independent, the dynamical invariant can be chosen to be time independent and to commute with the Hamiltonian. 
Because the invariant and the Hamiltonian then share a common time-independent eigenbasis, the conventional RF master equation can be written directly in the unified master-equation form of Eq.~\eqref{eqn:unifiedgenME}.

Second, when a time-independent rotating-frame Hamiltonian is unavailable, working in the time-dependent invariant eigenbasis $\ket{\mu_n(t)}$ allows us to derive a generalized RF master equation that preserves the essence of the secular approximation. As detailed in Appendix~\ref{sec:appendB}, the invariant eigenbasis allows us to construct the system propagator in Eq.~\eqref{eqn:freepropagator} required to switch to the interaction picture, thereby jointly fulfilling the roles of steps (a) and (b). Within the unified framework [Eq.~\eqref{eqn:unifiedgenME}]---which is defined under the Born--Markov approximation [step (c)] in the invariant eigenbasis---the IME dissipation coefficients [Eq.~\eqref{eqn:disscoefficient-IME}] constitute the most general form, as they retain all couplings without invoking the secular approximation in step (d).

Implementing step (d)---i.e., applying a secular approximation in the interaction picture---eliminates nonresonant (or nonsecular) coupling terms that are proportional to $e^{i[\theta_{m',n',\eta',\upsilon'}(t)-\theta_{m,n,\eta,\upsilon}(t)]}$ whenever
\begin{align}
    \theta_{m',n',\eta',\upsilon'}(t) -\theta_{m,n,\eta,\upsilon}(t)\neq 0\,.
    \label{eqn:generalRFsecularapproximation}
\end{align}
Specifically, enforcing $\theta_{m',n',\eta',\upsilon'}(t) =\theta_{m,n,\eta,\upsilon}(t)$ in the IME dissipation coefficients of Eq.~\eqref{eqn:disscoefficient-IME} reduces the IME dissipation coefficients to the generalized RF dissipation coefficients of Eq.~\eqref{eqn:disscoefficient-RF}. 

\paragraph*{Reduction to the conventional RF master equation.}
Since the conventional RF master equation is only defined if the system Hamiltonian can be made time independent in the rotating frame, it works only when the Rabi couplings are time independent, i.e., when $\Omega_{\eta,\upsilon}(t) = \Omega_{\eta,\upsilon}$. In this case, the dynamical invariant is time independent and has the same eigenbasis as the system Hamiltonian in the rotating frame. It follows that the complex transition amplitudes become time independent [$\xi_{m,n,\eta,\upsilon}(t) =\xi_{m,n,\eta,\upsilon}$] and that the instantaneous frequencies reduce to the effective driving frequencies for the transition [$\omega_{\vphantom{m',n'}m,n,\eta,\upsilon}(t,\tau) = \omega_{\vphantom{m',n'}m,n,\eta,\upsilon} = E_m-E_n- \omega^d_{\eta,\upsilon}$, where $E_m$ and $E_n$ are the eigenenergies of the Hamiltonian in the rotating frame]. Consequently, the total phase in Eq.~\eqref{eqn:phases} becomes linear in time, $\theta_{m,n,\eta,\upsilon}(t) = (E_m - E_n - \omega^d_{\eta,\upsilon})t$. This means that the secular approximation used to eliminate terms of the form given in Eq.~\eqref{eqn:generalRFsecularapproximation} neglects nonresonant (nonsecular) terms that oscillate in the interaction picture at frequencies
\begin{align}
    (E_{m'} - E_{n'} - E_m + E_n) - (\omega^d_{\eta',\upsilon'} - \omega^d_{\eta,\upsilon}) \neq 0\,.
    \label{eqn:conventionalRFnonsecular}
\end{align}
For terms where $\abs*{\omega^d_{\eta',\upsilon'} - \omega^d_{\eta,\upsilon}}$ is on the order of the drive frequencies, the rotating-wave approximation already ensures that this frequency difference is much larger than both $\abs*{E_{m'}-E_{n'}-E_m + E_n}$ and the decay rates. This means that neglecting these rapidly oscillating terms is equivalent to the rotating-wave approximation. In all other cases, additional secular approximations are required. In the limiting case where $\omega^d_{\eta',\upsilon'} - \omega^d_{\eta,\upsilon} = 0$ (such as for dephasing processes with $\eta=\upsilon$ and $\eta'=\upsilon'$, where $\omega^d_{\eta,\eta}=0$), Eq.~\eqref{eqn:conventionalRFnonsecular} reduces to $E_{m'} - E_{n'} - E_m + E_n \neq 0$. Neglecting nonresonant (nonsecular) terms that oscillate at $E_{m'} - E_{n'} - E_m + E_n \neq 0$ then requires that the eigenenergy splittings in the rotating frame are much larger than the slow dissipative timescale set by the maximum decay rate [step (d)],
\begin{align}
    \abs{E_{m'} - E_{n'}-E_m+E_n}\gg \text{max}\left(\abs{\Gamma_{m,n,m',n'}}\right)\,,
    \label{eqn:conventionalRFsecularapproximation}
\end{align}
which implies that the coupling strength must be much greater than the dissipation strength.

Under these conditions, the dissipation coefficients of Eq.~\eqref{eqn:disscoefficient-RF} reduce to the time-independent form
\begin{align}
    &\Gamma_{m,n,m',n'}\notag\\&= \left. \smashoperator{\sum\limits_{\substack{\eta\hphantom{'},\,\upsilon\hphantom{'}\\ \eta',\,\upsilon'}}} (\xi_{m',n',\eta',\upsilon'})^*\xi_{\vphantom{m',n'}m,n,\eta,\upsilon} \int\limits_0^{\infty}\!\!d\tau  e^{-i\omega_{\vphantom{m',n'}m,n,\eta,\upsilon}\tau} \bm{\Biggl(}\int\limits_0^{\infty}\!\!d\omega J_{\eta,\upsilon,\eta',\upsilon'}(\omega)\left\{e^{+i\omega \tau}n_b(\omega) + e^{-i\omega \tau}\left[1+n_b(\omega)\right]\right\}\bm{\Biggr)}\right\rvert_{\theta_{\vphantom{m',n'}m,n,\eta,\upsilon}\left(t\right)=\theta_{m',n',\eta',\upsilon'}\left(t\right)}.
\end{align}

\section{\label{sec:appendF} Dynamical invariant for an arbitrarily driven three-level system}
We consider the driven three-level $\Lambda$ system introduced in Sec.~\ref{sec:threelevel} and use the rotating-frame Hamiltonian given in Eq.~\eqref{eqn:3levelHamRF}. We express the Rabi couplings as $\Omega_{\eta,3}(t)=\abs*{\Omega_{\eta,3} (t)}e^{-i\phi_{\eta,3}}$ ($\eta = 1,2$), where $\abs*{\Omega_{\eta,3}(t)}$ is the time-dependent magnitude and $\phi_{\eta,3}$ is a time-independent phase. For the invariant construction below, we also allow the bare frequencies to depend on time, $\omega_\eta^0=\omega_\eta^0(t)$, subject to the equal-detuning condition $\Delta_1(t)=\Delta_2(t)=\Delta(t)$, where $\Delta_\eta(t)=\omega_\eta^0(t)-\omega_{\eta,3}^d$. The time dependence of the detunings is not used in the examples presented in the main text. 

To construct the dynamical invariant in the rotating frame, we introduce six real time-dependent auxiliary parameters $\theta(t),\beta(t), \phi(t),\nu(t),\eta(t)$, and $\zeta(t)$ to parametrize the orthonormal eigenbasis of the dynamical invariant:
\begin{subequations}
\label{eqn:eigenvectors}
\begin{align}
\ket{\mu_0(t)}
={}&\cos\beta(t)\cos\theta(t)\,\ket{1}
-\cos\beta(t)\sin\theta(t)\,e^{-i\nu(t)}\ket{2}
+\sin\beta(t)\,e^{-i(\eta(t)-\zeta(t))}\ket{3}\,,
\label{eqn:mu0}\\
\ket{\mu_+(t)}
={}&\left(\sin\phi(t)\sin\theta(t)
-\cos\phi(t)\sin\beta(t)\cos\theta(t)\,e^{-i\zeta(t)}\right)\ket{1}
\nonumber\\
&+\left(\sin\phi(t)\cos\theta(t)
+\cos\phi(t)\sin\beta(t)\sin\theta(t)\,e^{-i\zeta(t)}\right)e^{-i\nu(t)}\ket{2}
\nonumber\\
&+\cos\phi(t)\cos\beta(t)\,e^{-i\eta(t)}\ket{3}\,,
\label{eqn:muplus}\\
\ket{\mu_-(t)}
={}&\left(\cos\phi(t)\sin\theta(t)
+\sin\phi(t)\sin\beta(t)\cos\theta(t)\,e^{-i\zeta(t)}\right)\ket{1}
\nonumber\\
&+\left(\cos\phi(t)\cos\theta(t)
-\sin\phi(t)\sin\beta(t)\sin\theta(t)\,e^{-i\zeta(t)}\right)e^{-i\nu(t)}\ket{2}
\nonumber\\
&-\sin\phi(t)\cos\beta(t)\,e^{-i\eta(t)}\ket{3}\,.
\label{eqn:muminus}
\end{align}
\end{subequations}
The corresponding eigenvalues are $\lambda_0=0$, $\lambda_+=1$, and $\lambda_-=-1$. It can be checked explicitly that the states satisfy $\braket{\mu_j(t)}{\mu_k(t)}=\delta_{j,k}$. The invariant is then given by Eq.~\eqref{eqn:dynamicalInv}. The remaining task is to determine the six auxiliary parameters $\theta(t),\beta(t), \phi(t),\nu(t),\eta(t)$ and $\zeta(t)$. 

For $j\neq k$, Eq.~\eqref{eqn:invariantcondition} can be equivalently expressed as 
\begin{equation}
\bra{\mu_j(t)}\left(H^R_s(t)-i\dfrac{\partial}{\partial t}\right)\ket{\mu_k(t)}=0\,,
\label{eqn:projected_invariance}
\end{equation}
which is obtained using 
\begin{subequations}
\begin{align}
\bra{\mu_j(t)}\frac{\partial}{\partial t} I(t)\ket{\mu_k(t)}
=(\lambda_k-\lambda_j)\bra{\mu_j(t)}\frac{\partial}{\partial t} \ket{\mu_k(t)}
\end{align}
and
\begin{align}
\bra{\mu_j(t)}i[I(t),H^R_s(t)]\ket{\mu_k(t)}
=i(\lambda_j-\lambda_k)\bra{\mu_j(t)}H^R_s(t)\ket{\mu_k(t)}\,.
\end{align}
\end{subequations}
Writing out Eq.~\eqref{eqn:projected_invariance} and performing algebraic manipulations, we obtain the equations of motion for the auxiliary parameters: 
\begin{align}
\dfrac{d\theta(t)}{dt}&=-\dfrac{\tan\beta(t)}{2}\Big[
\abs{\Omega_{1,3}(t)}\sin\theta(t)\sin(\phi_{1,3}-\eta(t)+\zeta(t))+\abs{\Omega_{2,3}(t)}\cos\theta(t)\sin(\phi_{2,3}-\eta(t)+\nu(t)+\zeta(t))
\Big]\,,
\label{eqn:theta_ode}\\
\dfrac{d\nu(t)}{dt}&=-\frac{\tan\beta(t)}{2}\Bigg[
\frac{\abs{\Omega_{1,3}(t)}}{\cos\theta(t)}\cos(\phi_{1,3}-\eta(t)+\zeta(t))
+\frac{\abs{\Omega_{2,3}(t)}}{\sin\theta(t)}\cos(\phi_{2,3}-\eta(t)+\nu(t)+\zeta(t))
\Bigg]\,,
\label{eqn:xi_ode}\\
\dfrac{d\beta(t)}{dt}&=-\frac{\abs{\Omega_{1,3}(t)}}{2}\cos\theta(t)\sin(\phi_{1,3}-\eta(t)+\zeta(t))+\frac{\abs{\Omega_{2,3}(t)}}{2}\sin\theta(t)\sin(\phi_{2,3}-\eta(t)+\nu(t)+\zeta(t))\,,
\label{eqn:beta_ode}\\
\dfrac{d\phi(t)}{dt}&=\frac{1}{2\cos\beta(t)}\Big[
\abs{\Omega_{1,3}(t)}\sin\theta(t)\sin(\phi_{1,3}-\eta(t))
+\abs{\Omega_{2,3}(t)}\cos\theta(t)\sin(\phi_{2,3}-\eta(t)+\nu(t))
\Big]\,,
\label{eqn:varphi_ode}\\
\begin{split}
\dfrac{d\eta(t)}{dt}&=\Delta(t)
-\frac{\tan\beta(t)}{2}\Bigg[
2\abs{\Omega_{1,3}(t)}\cos\theta(t)\cos(\phi_{1,3}-\eta(t)+\zeta(t))
+\frac{\abs{\Omega_{2,3}(t)}\cos2\theta(t)}{\sin\theta(t)}
\cos(\phi_{2,3}-\eta(t)+\nu(t)+\zeta(t))
\Bigg]
\\
&\qquad\quad-\frac{\cot2\phi(t)}{\cos\beta(t)}\Big[
\abs{\Omega_{1,3}(t)}\sin\theta(t)\cos(\phi_{1,3}-\eta(t))
+\abs{\Omega_{2,3}(t)}\cos\theta(t)\cos(\phi_{2,3}-\eta(t)+\nu(t))
\Big]\,,
\end{split}
\label{eqn:eta_ode}\\
\intertext{and}
\begin{split}
\dfrac{d\zeta(t)}{dt}&=-\frac{\tan\beta(t)\cos2\theta(t)}{2}\Bigg[
\frac{\abs{\Omega_{1,3}(t)}}{\cos\theta(t)}\cos(\phi_{1,3}-\eta(t)+\zeta(t))
+\frac{\abs{\Omega_{2,3}(t)}}{\sin\theta(t)}\cos(\phi_{2,3}-\eta(t)+\nu(t)+\zeta(t))
\Bigg]
\\
&\quad-\frac{\cot\beta(t)}{2}\Big[
\abs{\Omega_{1,3}(t)}\cos\theta(t)\cos(\phi_{1,3}-\eta(t)+\zeta(t))
-\abs{\Omega_{2,3}(t)}\sin\theta(t)\cos(\phi_{2,3}-\eta(t)+\nu(t)+\zeta(t))
\Big]
\\
&\quad-\frac{\cot2\phi(t)}{\cos\beta(t)}\Big[
\abs{\Omega_{1,3}(t)}\sin\theta(t)\cos(\phi_{1,3}-\eta(t))
+\abs{\Omega_{2,3}(t)}\cos\theta(t)\cos(\phi_{2,3}-\eta(t)+\nu(t))
\Big]\,.
\end{split}
\label{eqn:zeta_ode}
\end{align}
Imposing the initial condition $\comm{I(0)}{H^R_s(0)}=0$ yields
\begin{subequations}
\begin{align}
\beta(0) &=  0\,, \\ 
\theta(0) &= \tan^{-1}\left[\frac{\abs{\Omega_{1,3}(0)}}{\abs{\Omega_{2,3}(0)}}\right]\,, \\ 
\nu(0) &= \phi_{1,3}-\phi_{2,3}\,, \\ 
\eta(0) &= \phi_{1,3}\,, 
\intertext{and}
\phi(0) &= \dfrac{1}{2}\tan^{-1}\left[\dfrac{\sqrt{\abs{\Omega_{1,3}(0)}^2 + \abs{\Omega_{2,3}(0)}^2}}{\Delta(0)}\right]\,.
\end{align}
\end{subequations}
Since there is a gauge freedom, $\zeta(0)$ can be chosen freely. We choose $\zeta(0)$ such that divergences at $t=0$ are avoided. Specifically, we choose $\zeta(0) = 0$ for Protocol 1, $\zeta(0) = -\pi/2$ for Protocol 2, and $\zeta(0) = -\pi/2$ for Protocol 3.

\section{\label{sec:appendG}Eliminating the time dependence of the generator}
For the periodic drive with constant Rabi couplings considered in Sec.~\ref{sec:spectrum}, the Hamiltonian in the rotating frame and the dynamical invariant are time independent. The dissipation coefficients of the LF and IME are, however, in general, time dependent. This appendix describes the approximations that allow us to remove the time dependence of the dissipation coefficients of the IME. The dissipation coefficients of the LF master equation can be treated analogously. The time-independent dissipation coefficients yield the generator that is used for the steady-state spectral calculations.

For constant Rabi couplings, we have $\varsigma_{m,n,\eta,\upsilon}(t) =-\omega^d_{\eta,\upsilon}t$ and $\xi_{m,n,\eta,\upsilon}(t) = \xi_{m,n,\eta,\upsilon}$. We neglect the Lamb shifts [i.e., we drop the imaginary parts of $\Gamma_{m,n,m',n'}(t)$] and define $\gamma_{m,n,m',n'}(t)=\Re\left[\Gamma_{m,n,m',n'}(t)\right]$. Lamb shifts introduce small quantitative shifts that are not expected to affect the qualitative behavior of the fluorescence spectra and are neglected in Sec.~\ref{sec:spectrum}. Using these simplifications in Eq.~\eqref{eqn:disscoefficient-IME}, we find
\begin{align}
    \gamma_{m,n,m',n'}(t)
    ={}&\Re\left\{\!
    \sum\limits_{\substack{\eta\hphantom{'},\,\upsilon\hphantom{'}\\
                          \eta',\,\upsilon'}}\!\!
    e^{-i(\omega^d_{\eta,\upsilon}
           -\omega^d_{\eta',\upsilon'})t}
    (\xi_{m',n',\eta',\upsilon'})^*
    \xi_{\vphantom{m',n'}m,n,\eta,\upsilon}
    \left[
        R_{\eta,\upsilon,\eta',\upsilon'}
        (E_n-E_m+\omega^d_{\eta,\upsilon})
        +iI_{\eta,\upsilon,\eta',\upsilon'}
        (E_n-E_m+\omega^d_{\eta,\upsilon})
    \right]\!
    \right\},
    \label{eqn:time_dependent_dissipation}
\end{align}
where, for the zero-temperature bath considered here,
\begin{subequations}
\label{eqn:bath_response_parts}
\begin{align}
    R_{\eta,\upsilon,\eta',\upsilon'}(\omega_0)
    &=\Re\left\{
        \int_0^\infty d\omega\,
        J_{\eta,\upsilon,\eta',\upsilon'}(\omega)
        \int_0^\infty d\tau\,
        e^{-i(\omega-\omega_0)\tau}
    \right\}
\end{align}
and
\begin{align}
    I_{\eta,\upsilon,\eta',\upsilon'}(\omega_0)
    &=\Im\left\{
        \int_0^\infty d\omega\,
        J_{\eta,\upsilon,\eta',\upsilon'}(\omega)
        \int_0^\infty d\tau\,
        e^{-i(\omega-\omega_0)\tau}
    \right\}.
\end{align}
\end{subequations}
The functions $R$ and $I$ are the real and imaginary parts of the integrated bath response, respectively. We use the abbreviation $R_{\eta,\upsilon}=R_{\eta,\upsilon,\eta,\upsilon}$. The explicit time dependence in Eq.~\eqref{eqn:time_dependent_dissipation} arises from the differences between the driving frequencies. For the diagonal couplings, we have $\omega^d_{\eta,\eta}=0$. For the nondiagonal couplings, we have $\omega^d_{1,2}=\omega^d_{1,3}-\omega^d_{2,3}$, $\omega^d_{2,1}=-\omega^d_{1,2}$, and $\omega^d_{3,\eta}=-\omega^d_{\eta,3}$.

\paragraph*{Secular approximation for driving frequencies.}
For the $\Lambda$ system of interest in this work, the driving frequencies $\omega^d_{1,3}$ and $\omega^d_{2,3}$ are much larger than the other relevant dynamical and dissipative scales. In particular,
\begin{align}
    \abs{\omega^d_{1,2}},\abs{\omega^d_{2,1}}
    \ll
    \abs{\omega^d_{1,3}},\abs{\omega^d_{3,1}},
    \abs{\omega^d_{2,3}},\abs{\omega^d_{3,2}}.
\end{align} 
Terms such as $e^{\pm i(\omega^d_{1,3}\pm \omega^d_{\eta,\eta})t}$, $e^{\pm i(\omega^d_{2,3}\pm \omega^d_{\eta,\eta})t}$, $e^{\pm i(\omega^d_{1,3}-\omega^d_{3,1})t}$, and $e^{\pm i(\omega^d_{2,3}-\omega^d_{3,2})t}$ therefore oscillate rapidly on the system's dynamical timescales and are neglected under the secular approximation. Formally, this approximation requires
\begin{align}
    \abs{\omega^d_{\eta,\upsilon}-\omega^d_{\eta',\upsilon'}}
    \gg \text{max}\abs{\gamma_{m,n,m',n'}(t)}.
    \label{eqn:optical_secular_condition}
\end{align}
This does not remove terms that oscillate at frequencies $\omega^d_{1,2}$ or $2\omega^d_{1,2}$, which are comparable to other energy scales for a $\Lambda$ system. Applying Eq.~\eqref{eqn:optical_secular_condition}, the remaining coefficients read
\begin{align}
    \gamma_{m,n,m',n'}(t)
    ={}&\sum_{\eta=1}^{2}
    (\xi_{m',n',\eta,3})^*\xi_{m,n,\eta,3}
    R_{\eta,3}(E_n-E_m+\omega^d_{\eta,3})
    \nonumber\\
    &+
    (\xi_{m',n',1,2})^*\xi_{m,n,1,2}
    R_{1,2}(E_n-E_m+\omega^d_{1,2})
    \nonumber\\
    &+
    (\xi_{m',n',2,1})^*\xi_{m,n,2,1}
    R_{2,1}(E_n-E_m-\omega^d_{1,2})
    \nonumber\\
    &+\sum_{\eta,\eta'=1}^{3}
    (\xi_{m',n',\eta',\eta'})^*\xi_{m,n,\eta,\eta}
    R_{\eta,\eta,\eta',\eta'}(E_n-E_m)
    \nonumber\\
    &+\Re\Big\{
    e^{-i\omega^d_{1,2}t}
    (\xi_{m',n',2,3})^*\xi_{m,n,1,3}
    \big[R_{1,3,2,3}(E_n-E_m+\omega^d_{1,3})
    +iI_{1,3,2,3}(E_n-E_m+\omega^d_{1,3})\big]\Big\}
    \nonumber\\
    &+\Re\Big\{
    e^{+i\omega^d_{1,2}t}
    (\xi_{m',n',1,3})^*\xi_{m,n,2,3}
    \big[R_{2,3,1,3}(E_n-E_m+\omega^d_{2,3})
    +iI_{2,3,1,3}(E_n-E_m+\omega^d_{2,3})\big]\Big\}
    \nonumber\\
    &+\Re\Big\{
    e^{+i\omega^d_{1,2}t}
    (\xi_{m',n',3,2})^*\xi_{m,n,3,1}
    iI_{3,1,3,2}(E_n-E_m-\omega^d_{1,3})\Big\}
    \nonumber\\
    &+\Re\Big\{
    e^{-i\omega^d_{1,2}t}
    (\xi_{m',n',3,1})^*\xi_{m,n,3,2}
    iI_{3,2,3,1}(E_n-E_m-\omega^d_{2,3})\Big\}
    \nonumber\\
    &+\Re\Big\{
    e^{-2i\omega^d_{1,2}t}
    (\xi_{m',n',2,1})^*\xi_{m,n,1,2}
    \big[R_{1,2,2,1}(E_n-E_m+\omega^d_{1,2})
    +iI_{1,2,2,1}(E_n-E_m+\omega^d_{1,2})\big]\Big\}
    \nonumber\\
    &+\Re\Big\{
    e^{+2i\omega^d_{1,2}t}
    (\xi_{m',n',1,2})^*\xi_{m,n,2,1}
    \big[R_{2,1,1,2}(E_n-E_m-\omega^d_{1,2})
    +iI_{2,1,1,2}(E_n-E_m-\omega^d_{1,2})\big]\Big\}
    \nonumber\\
    &+\Re\Big\{
    e^{-i\omega^d_{1,2}t}\sum_{\eta=1}^{3}
    (\xi_{m',n',\eta,\eta})^*\xi_{m,n,1,2}
    \big[R_{1,2,\eta,\eta}(E_n-E_m+\omega^d_{1,2})
    +iI_{1,2,\eta,\eta}(E_n-E_m+\omega^d_{1,2})\big]\Big\}
    \nonumber\\
    &+\Re\Big\{
    e^{+i\omega^d_{1,2}t}\sum_{\eta=1}^{3}
    (\xi_{m',n',\eta,\eta})^*\xi_{m,n,2,1}
    \big[R_{2,1,\eta,\eta}(E_n-E_m-\omega^d_{1,2})
    +iI_{2,1,\eta,\eta}(E_n-E_m-\omega^d_{1,2})\big]\Big\}
    \nonumber\\
    &+\Re\Big\{
    e^{+i\omega^d_{1,2}t}\sum_{\eta=1}^{3}
    (\xi_{m',n',1,2})^*\xi_{m,n,\eta,\eta}
    \big[R_{\eta,\eta,1,2}(E_n-E_m)
    +iI_{\eta,\eta,1,2}(E_n-E_m)\big]\Big\}
    \nonumber\\
    &+\Re\Big\{
    e^{-i\omega^d_{1,2}t}\sum_{\eta=1}^{3}
    (\xi_{m',n',2,1})^*\xi_{m,n,\eta,\eta}
    \big[R_{\eta,\eta,2,1}(E_n-E_m)
    +iI_{\eta,\eta,2,1}(E_n-E_m)\big]\Big\}.
    \label{eqn:remaining_time_dependence}
\end{align}

\paragraph*{Removing the remaining time dependence.}
To remove the remaining time dependence, we switch off the $(1{\leftrightarrow}2)$ decay/excitation channel by setting $g_{1,2,k}=0$ for all bath modes; we hence have $g_{2,1,k}=0$ and $\Gamma_{1,2}=\Gamma_{2,1}=0$. This removes all time-dependent contributions due to cross couplings involving the $(1\leftrightarrow2)$ channels. Only the four explicitly time-dependent contributions that couple the $(1{\leftrightarrow}3)$ and $(2{\leftrightarrow}3)$ channels remain. These cross-dissipative terms have a rate that is proportional to $\sqrt{\Gamma_{1,3}\Gamma_{2,3}}$. Their phase factors oscillate at $\pm(\omega^d_{1,3}-\omega^d_{2,3})$. These contributions can be neglected provided
\begin{align}
    \omega^d_{1,2}
    \gg
       \sqrt{\Gamma_{1,3}\Gamma_{2,3}}\,.
    \label{eqn:remaining_optical_cross_condition}
\end{align}
Applying Eq.~\eqref{eqn:remaining_optical_cross_condition}, the resulting time-independent IME coefficients that are accounted for in our spectral calculations are
\begin{align}
    \gamma_{m,n,m',n'}
    ={}&
    (\xi_{m',n',1,3})^*\xi_{m,n,1,3}
    R_{1,3}(E_n-E_m+\omega^d_{1,3})
   +
    (\xi_{m',n',2,3})^*\xi_{m,n,2,3}
    R_{2,3}(E_n-E_m+\omega^d_{2,3})
    \nonumber\\
    &+\sum_{\eta,\eta'=1}^{3}
    (\xi_{m',n',\eta',\eta'})^*\xi_{m,n,\eta,\eta}
    R_{\eta,\eta,\eta',\eta'}(E_n-E_m).
    \label{eqn:eq_ch3_diss_inv}
\end{align}
With this, the master equation becomes
\begin{align}
    \frac{d\rho(t)}{dt}
    =-i[H_s^R,\rho(t)]
    +\sum_{m,n,m',n'}\gamma_{m,n,m',n'}
    \Big[
        F_{m,n}\rho(t) F_{m',n'}^{\Dag}
        +F_{m',n'}\rho(t) F_{m,n}^{\Dag}
        -F_{m',n'}^{\Dag}F_{m,n}\rho(t)
        -\rho(t) F_{m,n}^{\Dag}F_{m',n'}
    \Big].
    \label{eqn:eq_ME}
\end{align}
The sum in Eq.~\eqref{eqn:eq_ch3_diss_inv} contains all nine longitudinal contributions, including correlations between different diagonal couplings. These terms are time independent because the diagonal couplings have zero driving frequency. This implies that the time-independent generator accounts for the correlations between the different longitudinal couplings associated with dephasing processes. We emphasize that the sums over $m,n,m',n'$ in Eq.~\eqref{eqn:eq_ME} are unrestricted, since no further secular approximation that would remove transitions associated with different invariant-basis states was imposed to obtain Eqs.~\eqref{eqn:eq_ch3_diss_inv} and \eqref{eqn:eq_ME}.
\twocolumngrid
\bibliography{references}
\end{document}